\documentclass[journal=jacsat, manuscript=article]{achemso}

\usepackage[version=3]{mhchem}
\usepackage[utf8]{inputenc}
\usepackage{graphicx}
\usepackage{wasysym}
\usepackage{natbib}
\usepackage{amsmath}
\usepackage{amsfonts}
\usepackage{amssymb}
\usepackage{xcolor}
\usepackage{soul}
\usepackage{setspace}
\usepackage{dcolumn}
\usepackage{color}
\usepackage{bm}
\usepackage{graphicx,wrapfig,lipsum}
\usepackage[normalem]{ulem}
\usepackage{pdfpages}

\usepackage[labelfont=bf]{caption}

\usepackage{url}

\usepackage{url}
\usepackage[ colorlinks = true,
linkcolor = red,
urlcolor  = red,
citecolor = blue,
anchorcolor = red]{hyperref}
\usepackage{bibentry}
\usepackage{fancyhdr}
\author{Ranit Dutta}
\affiliation{These authors contributed equally to this work}
\alsoaffiliation[University]{Department of Physics, Indian Institute of Science, Bangalore 560012, India}

\author{Ayan Ghosh}
\affiliation{These authors contributed equally to this work}
\alsoaffiliation[University]{Department of Physics, Indian Institute of Science, Bangalore 560012, India}

\author{Shinjan Mandal}
\affiliation[University]{Department of Physics, Indian Institute of Science, Bangalore 560012, India}

\author{K. Watanabe}
\affiliation[University]{Research Center for Functional Materials, National Institute for Materials Science, 1-1 Namiki, Tsukuba 305-0044, Japan}
\author{T. Taniguchi}
\affiliation[University]{International Center for Material Nanoarchitectonics, National Institute for Materials Science,  1-1 Namiki, Tsukuba 305-0044, Japan}

\author{H.R. Krishnamurthy}
\affiliation[University]{Department of Physics, Indian Institute of Science, Bangalore 560012, India}

\author{Sumilan Banerjee}
\affiliation[University]{Department of Physics, Indian Institute of Science, Bangalore 560012, India}

\author{Manish Jain}
\affiliation[University]{Department of Physics, Indian Institute of Science, Bangalore 560012, India}

\author{Anindya Das}
\email{anindya@iisc.ac.in}
\affiliation[University]{Department of Physics, Indian Institute of Science, Bangalore 560012, India}

\title{Electric field tunable superconductivity with competing orders in twisted bilayer graphene near magic-angle}

\begin{document}

 \begin{abstract}
    \begin{spacing}{2}
    
    {\bf Superconductivity (SC) in twisted bilayer graphene (tBLG) has been explored by varying carrier concentrations, twist angles, and screening strength, with the aim of uncovering its origin and possible connections to strong electronic correlations in narrow bands and various resulting broken symmetries. However, the link between the tBLG band structure and the onset of SC and other orders largely remains unclear. In this study, we address this crucial gap by examining in-situ band structure tuning of a near magic-angle ($\theta \approx0.95^\circ$) tBLG device with displacement field ($D$) and reveal remarkable competition between SC and other broken symmetries. At zero $D$, the device exhibits superconducting signatures without the resistance peak at half-filling, a characteristic signature with a strong electronic correlation. As $D$ increases, the SC is suppressed, accompanied by the appearance of a resistance peak at half-filling. Hall density measurements reveal that at zero $D$, SC arises around the van Hove singularity (vHs) from an isospin or spin-valley unpolarized band. At higher $D$, the suppression of SC coincides with broken isospin symmetry near half-filling with lifted degeneracy ($g_d \sim 2$). Additionally, as the SC phase becomes weaker with $D$, vHs shifts to higher fillings, highlighting the modification of the underlying band structure with the applied electric field. These findings, with recent theoretical study on SC in tBLG, highlight the competition, rather being connected concomitantly, between SC and other orders promoted by broken symmetries}.
\end{spacing}
    \end{abstract}
    
\begin{spacing}{2}
 A relative twist angle between two or more van der Waals layers\cite{cao2018unconventional1,lu2019superconductors,cao2018correlated1,liu2020tunable,cao2020tunable,su2023superconductivity,kuiri2022spontaneous,park2021tunable,hao2021electric,park2022robust,zhang2022promotion}, leading to the formation of moir\'e superlattice, 
 has opened up a completely new field of condensed matter research known as `twistronics'. Observations of interaction-driven emergent phenomena like superconductivity (SC) \cite{cao2018unconventional1,yankowitz2019tuning,lu2019superconductors,su2023superconductivity,doi:10.1126/science.abc2836,paul2022interaction,codecido2019correlated,arora2020superconductivity,stepanov2020untying,saito2020independent,liu2021tuning,tian2023evidence,PhysRevX.8.031089,park2021tunable,hao2021electric,park2022robust,zhang2022promotion}, correlated Mott insulators \cite{cao2018correlated1,lu2019superconductors}, orbital ferromagnetism \cite{sharpe2019emergent}, anomalous Hall \cite{tseng2022anomalous} and quantized anomalous Hall effect \cite{Serlin900},  nematicity \cite{doi:10.1126/science.abc2836}, 
 Chern insulators \cite{saito2021hofstadter,wu2021chern,das2021symmetry,chen2020tunable,pierce2021unconventional}, strange metal \cite{cao2020strange}, giant thermopower \cite{paul2022interaction} and breakdown of semi-classical description of Mott's formula \cite{paul2022interaction,ghawri2022breakdown,ghosh2024evidence} are 
 ubiquitous in this area. Among the family of twisted heterostructures, magic-angle twisted bilayer graphene (MAtBLG)~\cite{cao2018unconventional1,lu2019superconductors,cao2018correlated1,cao2021pauli,yankowitz2019tuning,lin2022spin,doi:10.1126/science.abc2836,wu2021chern,das2021symmetry,cao2020strange,stepanov2020untying,liu2021tuning,zondiner2020cascade,wong2020cascade} with a twist angle of $\theta_M\simeq1.05^\circ-1.1^\circ$ between two sheets of graphene monolayer is being extensively studied for hosting flat bands~\cite{bistritzer2011moire} and having a rich phase diagram with such strongly correlated phases. Inter-layer hybridization between the rotated layers facilitates the formation of these isolated flat bands which 
 causes the effective electronic kinetic energy to become much smaller than the effective 
 coulomb interactions~\cite{cao2018correlated1} leading to the realization of such correlated phases. The appearance of correlated insulators (CI)~\cite{cao2018correlated1,lu2019superconductors} at integer  moir\'e filling $\nu$ and 
 a superconducting 
 dome when the CIs are doped 
 away from the integer filling~\cite{cao2018unconventional1,lu2019superconductors} shows a strong resemblance to the behavior of 
 high-$T_c$ ($>100$ K) cuprate superconductors ~\cite{xu_vortex-like_2000}. 

 Recent studies have, however, raised questions as to whether and to what extent superconductivity in tBLG is intimately connected with correlation effects~\cite{stepanov2020untying,saito2020independent,arora2020superconductivity}. In these reports, superconductivity has been observed even in the absence of correlated insulator states 
 by varying the dielectric thickness to control the coulomb screening~\cite{stepanov2020untying}, and even at angles detuned away from $\theta_M$, with the SC phase sometimes taking over the whole of the phase space without any signature of correlated states~\cite {stepanov2020untying,saito2020independent,arora2020superconductivity}. The decoupling of superconductivity and correlated insulator behaviour 
 in the presence of screened coulomb interactions~\cite{liu2021tuning} or the increased bandwidth of the flat bands (angles away from $\theta_M\pm0.05^o$ ) indicate that these two phenomena may not be always intimately connected, and may even be 
 competing\cite{stepanov2020untying,saito2020independent} and electron-phonon mechanism may be playing a significant role in the origin of superconductivity in tBLG\cite{stepanov2020untying,lian2019twisted,cea2021coulomb,shavit2021theory}. Thus, unlike other graphene-based systems such as Bernal bilayer graphene (BBG)\cite{zhou2022isospin,zhang2023enhanced,li2024tunable}, rhombohedral trilayer graphene (RTG)\cite{zhou2021superconductivity}, ABC-trilayer graphene/aligned with hBN\cite{chen2019signatures}, alternatingly twisted trilayer\cite{park2021tunable,hao2021electric}, and multilayer graphene\cite{park2022robust,zhang2022promotion}, where superconductivity consistently emerges close to other symmetry-broken phases, it remains unclear 
 whether superconductivity and other symmetry-broken phases are fundamentally linked in twisted bilayer graphene (tBLG). 
\begin{figure*}[htbp]
\begin{center}
    \centerline{\includegraphics[width=0.95\textwidth]{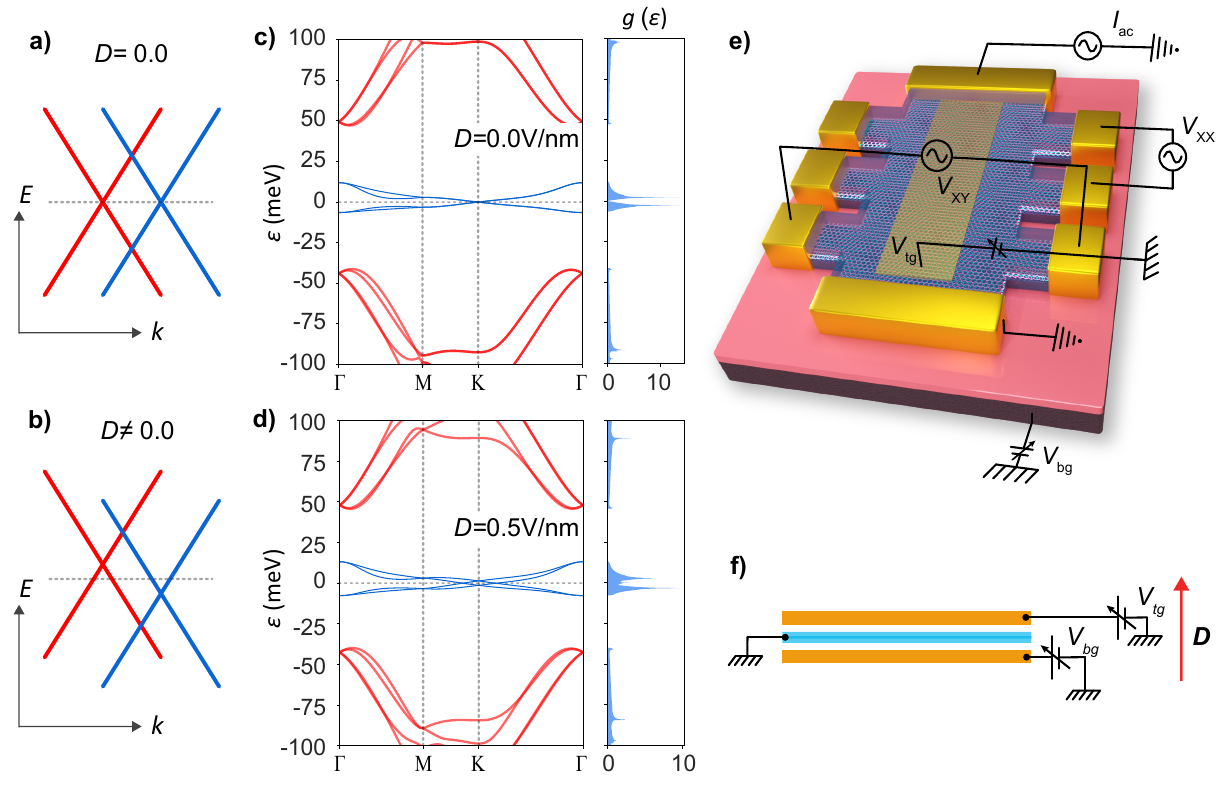}}
\end{center}
\caption{\label{mf1}\textbf{Band structure and device schematic.} Schematic presentation of band crossing of top and bottom layer graphene at zero $D$, \textbf{a)} and finite $D$, \textbf{b)}. The calculated band structures using a tight-binding model on relaxed structures of twist angle, $\theta\sim 0.95^\circ$ tBLG for $D = 0$, \textbf{c)} and $D = 0.5$V/nm, \textbf{d)} (For the theoretical calculation of $D$ see Methods and \textcolor{purple}{Supplementary Information SI-12 (II)}). \textbf{e)} Device set-up for measuring $R_{xx}$ and $R_{xy}$. Using the combination of a metal top gate  ($V_{tg}$) and a global SiO$_2/$Si back gate ($V_{bg}$), the carrier density, $n$, and the perpendicular displacement field, $D$, were controlled independently. \textbf{f)} The schematic for the positive displacement field when $V_{bg}$ is positive but $V_{tg}$ is negative. All the measurements are done with constant current bias and using a standard low-frequency lock-in technique at frequency $f = 13$  Hz in a cryo-free dilution refrigerator.}
\label{Figure1}
\end{figure*} 

To answer these questions, one needs to investigate specific features of tBLG, other than electron-electron and electron-phonon interaction, e.g., band structure details such as van Hove singularity (vHs) that can crucially influence superconductivity. However, the exploration of how the appearance of superconductivity is connected to the features
of such band structure details in tBLG is missing in the literature. Here, we address this issue through the in-situ tuning of the band structure by a displacement field that continuously tunes a vHs further away from half-filling and thus induces other symmetry-breaking orders, presumably competing with SC. As shown schematically in Figure.~\ref{mf1}a-d, applying a perpendicular electric field can tune the band structure of a tBLG with a twist angle slightly away from the magic angle, and can give insights into the intimate connection between the 
band structure and SC. Thus, we present 
a comprehensive study on tuning the SC phase of a near magic-angle tBLG ($\sim 0.95^\circ$) device as a function of a perpendicular electric field (displacement field, $D$). At zero displacement field ($D = 0.00$ V/nm), the device shows resistance peaks at the Dirac point (CNP) and full band filling ($\nu = \pm 4$) also known as superlattice gaps, but without any signature of resistance peaks at 
half-filling ($\nu \sim\pm 2$). Most interestingly, the device shows the signature of SC with resistance dropping to almost zero with decreasing 
temperature away from 
half-filling ($\nu \sim +2.7$). Typical behaviours such as 
a superconducting dome with temperature ($T$), non-linear current-voltage ($I-V$) characteristics, and suppression of the superconducting phase with a magnetic field ($B_\perp$) establish the robustness of the SC phase of the device at $D = 0.00$ V/nm. With increasing $D$, the SC phase is suppressed with the appearance of the resistance peak at the half-filling ($\nu \sim +2$). The tuning of the SC is further quantified from the $D$ dependent critical temperature, $T_c$, critical current, $I_c$, and critical magnetic field, $B_c$, (both parallel and perpendicular). To understand the role of the band structure, we study the Hall density (Hall filling$/$Normalized Hall density, $\nu_H$) as a function of the filling ($\nu$) at different $D$, which reveals the following: (i)  at  $D = 0.00$ V/nm, the SC arises precisely around the van Hove singularity (vHs) from an isospin-unpolarized band with degeneracy, $g_d = 4$, (ii) at higher $D$ magnitudes, the suppression of SC is concomitant with the broken isospin symmetry near half-filling, with lifted degeneracy of $g_d \sim 2$. 

\begin{figure*}[htbp]
\begin{center}
    \centerline{\includegraphics[width=0.65\textwidth]{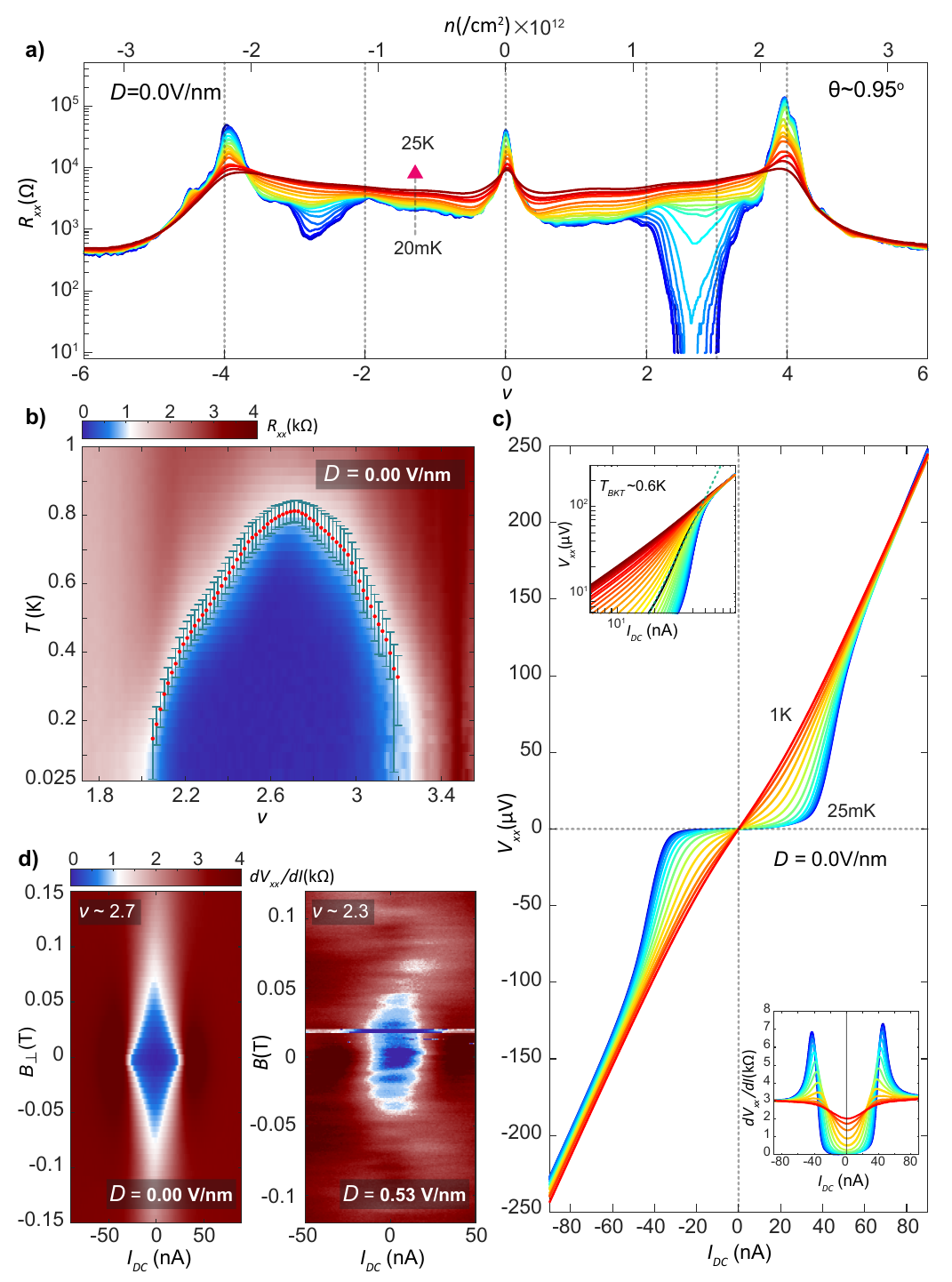}}
\end{center}
\caption{\label{mf2}\textbf{Robust superconductivity at zero $D$.} \textbf{a)} Measured $R_{xx}$ versus carrier filling, $\nu$, from $T = 20$ mK up to $25$ K at $D = 0$ V/nm. The resistance peaks at the Dirac point, and full band filling ($\nu = +4, -4$) are seen without any signature of resistance peaks at other integer fillings. 
The $R_{xx}$ value approaches close to zero on the electron side at $T = 20$ mK between $\nu \sim 2 - 3$ marking the superconducting
region. 
\textbf{b)} 2-d colormap of $R_{xx}$ as a function of $\nu$ and $T$ showing the superconductivity dome around optimal doping, $\nu_c \sim 2.6-2.7$. The red dots are the critical temperatures, $T_c$, corresponding to the $ 50\%$ value of normal state resistance, $R_n$. The error bars refer to the corresponding temperatures, $T$, at which$R_{xx}$ drops to $60\%$ and $ 40\%$ of $R_n$. \textbf{c)} $I-V$ characteristics of SC region at $\nu \approx +2.7$ with variation of $T$ from $25$ mK to $1$ K. 
The lower inset shows the $dV_{xx}/dI$ versus $I_{DC}$. 
In the upper inset, $V_{xx} \propto I^{3}$ behaviour is seen suggesting a typical $2$D superconductor with a Berezinskii-Kosterlitz-Thouless transition at $T_{BKT} \sim 600$ mK. \textbf{d)} $dV_{xx}/dI$ versus $I_{DC}$ and $B_\perp$ for $\nu,D = +2.7, 0.00$ V/nm (left panel) and $+2.3, 0.50$ V/nm (right panel). The Fraunhofer-like oscillations can be seen when the SC is suppressed at higher $D$.} 
\label{Figure2}
\end{figure*}
 \end{spacing}
\section*{Results}
 \subsection*{Device set-up and characterization}
 \begin{spacing}{2}
For our study, we have fabricated a dual-gated hBN encapsulated twisted bilayer graphene device using the modified `cut and stack' technique from an exfoliated single monolayer graphene sheet on a SiO$_2/$Si substrate~\cite{ghosh2023evidence}. Fabrication details are mentioned in the \textcolor{purple}{Supplementary Information SI- 1}. A metal layer above the top hBN acts as the top gate of the device, whereas the Silicone (Si) layer acts as the global back gate. The dual-gated device structure helps to tune the total number density, $n$, and the perpendicular displacement field, $D$, independently ~\cite{liu2020tunable,cao2020tunable,su2023superconductivity,kuiri2022spontaneous,park2021tunable,hao2021electric} in our tBLG device, as shown schematically in Figure.~\ref{mf1}f (see \textcolor{purple}{SI- 2} for details). Figure.~\ref{mf1}e shows the device schematic together with the measurement scheme to measure longitudinal ($R_{xx}$) and transverse ($R_{xy}$) resistance in a standard Hall-bar geometry set-up. Figure.~\ref{mf2}a  shows the longitudinal resistance, $R_{xx}$, versus the filling factor, $\nu$, with increasing temperature from the base temperature of $T=20$ mK up to $25$ K at $D = 0.00$ V/nm. The carrier density, $n$, measured relative to the charge neutrality point (CNP) 
is converted to moir\'e filling factor, $\nu=\frac{4n}{n_s}$, where $n_s$ is the requisite carrier density to fill$/$empty the low energy flat bands (i.e $4$ electron$/$holes per moir\'e superlattice unit cell)~\cite{cao2018unconventional1,lu2019superconductors}. High insulating resistance peaks 
at moir\'e integer filling, $\nu=\pm4$, on either side of the CNP indicates the presence of a single-particle band gap above and below the low energy flat bands. The twist angle from the carrier densities for $\nu=\pm4$ corresponds to  
 $\theta\approx0.95^\circ\pm0.02^\circ$ for our device, which is around $13\%$ smaller than the magic angle $\theta_M\approx1.1^0$. 
 Unlike other tBLG devices closer to the magic-angle~\cite{cao2018unconventional1,lu2019superconductors,cao2018correlated1,yankowitz2019tuning,stepanov2020untying,saito2020independent}, we do not observe any resistance peaks 
 corresponding to the correlated insulators at half-filling, $\nu=\pm2$, or at other integer fillings\cite{lu2019superconductors}. All the measurements are done using the standard lock-in technique at $f\sim 13 $ Hz in a cryo-free dilution fridge for several thermal cycles.
\end{spacing}

\subsection*{Superconductivity at zero $D$}
\begin{spacing}{2}
 As seen in Figure.~\ref{mf2}a, we see a clear signature of superconductivity on the electron side with the $R_{xx}$ value going down to almost zero between $\nu=+2.2$ and $\nu=+3$ at the base temperature of $T=20$ mK. Though the signature of resistance decrement with lowering of the temperature on the hole side between $\nu=-2.2$ and $\nu=-3$ can be seen in Figure.~\ref{mf2}a, 
 in this region the $R_{xx}$ value remains finite at the lowest measured temperature ($\sim20 $ mK). From here on, we will only focus on the superconducting region on the electron side ($+2<\nu<+3$). Figure.~\ref{mf2}b shows $R_{xx}(\nu,T)$ in the $2-$d colormap. A superconducting dome around $\nu_c \sim 2.6-2.7$ (optimal doping) is visible clearly. The temperature dependence of $V_{xx}$ versus $I_{DC}$ traces are shown in Figure.~\ref{mf2}c, which is obtained by integrating $dV_{xx}/dI$ as a function of $I_{DC}$ as shown in the lower inset of Figure.~\ref{mf2}c. The traces in Figure.~\ref{mf2}c show the classic evolution of $2$D superconductors from step-like transitions at low temperature to ohmic dependence around $1$ K. The upper inset in Figure.~\ref{mf2}c shows the BKT (Berezinskii-Kosterlitz-Thouless) transition, where the evolution of the $V_{xx} \sim I^{\alpha}_{DC}$ power law near the critical bias current gives the BKT transition temperature, $T_{BKT} \sim 0.6$ K corresponding to $V_{xx} \sim I_{DC}^{3}$. 
 Figure.~\ref{mf2}b and~\ref{mf2}c establish the robustness of the superconductivity in our device at $D = 0.00$ V/nm. Further, the superconductivity is killed by applying a perpendicular magnetic field ($B_{c\perp} \approx 60-70 $ mT) as shown in Figure.~\ref{mf2}d left panel. Though the Fraunhofer-like oscillation pattern~\cite{cao2018unconventional1,lu2019superconductors,yankowitz2019tuning,paul2022interaction,su2023superconductivity,arora2020superconductivity,park2021tunable} is not seen at  $D=0.00$ V/nm, it is visible when the superconductivity gets suppressed at finite $D$, as shown in the right panel of Figure. \ref{mf2}d (details in \textcolor{purple}{SI- 4}). The absence of the Fraunhofer pattern suggests the existence of a more homogeneous SC phase \cite{zhou2024double}. These observations critically establish the occurrence of a robust $2$D superconductivity \cite{cao2018unconventional1,lu2019superconductors,stepanov2020untying,saito2020independent,arora2020superconductivity,hao2021electric,park2021tunable} seen here in the near magic-angle tBLG at zero $D$.
\end{spacing}

\begin{figure*}[htbp]
\begin{center}
    \centerline{\includegraphics[width=1.0\textwidth]{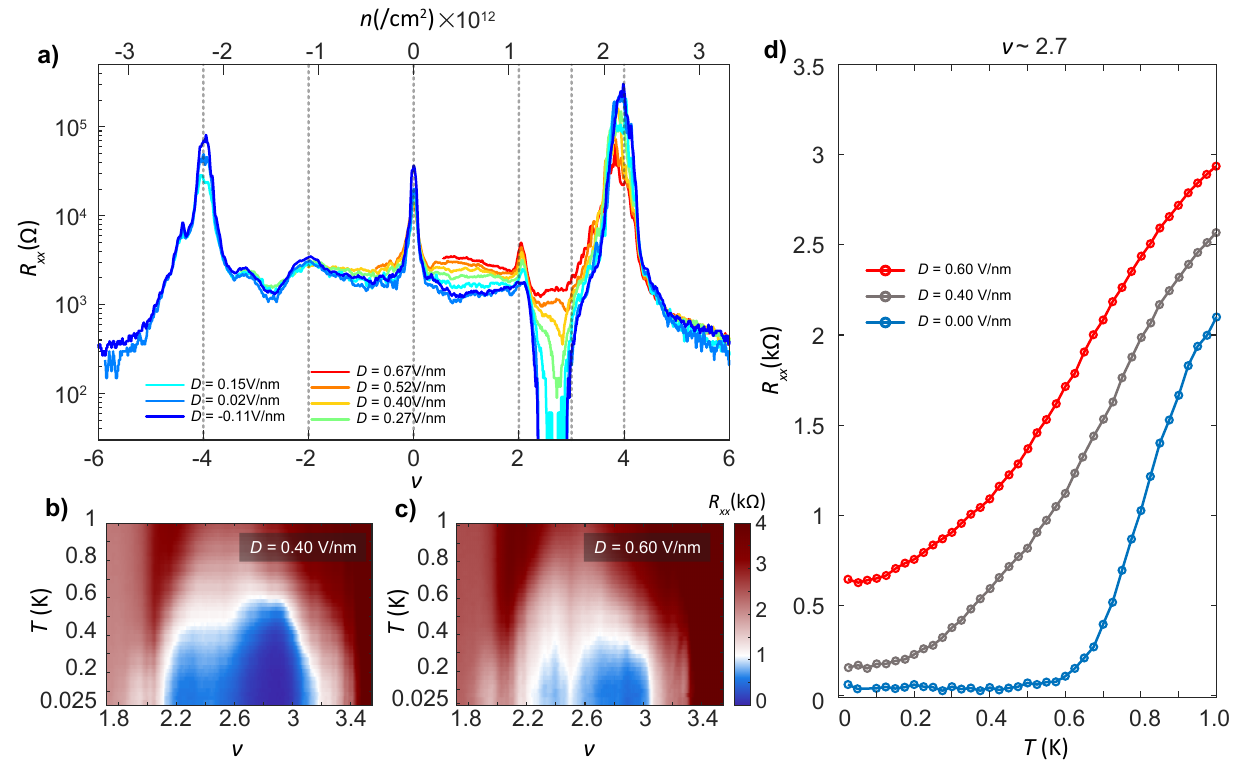}}
\end{center}
\caption{\label{mf3}\textbf{Tunability of superconductivity with $D$.} \textbf{a)} $R_{xx}$ versus $\nu$ at the base temperature 
for different $D$. The SC weakens with increasing $D$, and a resistance peak emerges at $\nu \sim 2$ with a background resistance increment around $\nu \sim 1$. 
\textbf{b)} and \textbf{c)} represent the superconducting domes of $R_{xx}(\nu,T$) for $D = 0.40$ V/nm and $ 0.60$ V/nm, respectively. The superconducting domes' width and height decrease with increasing $D$ compared to the zero $D$ case in Figure.~\ref{mf2}b. 
\textbf{d)} Plots of $R_{xx}$ vs $T$ for different $D$ fields at $\nu \sim 2.7$.}
\label{Figure3}
\end{figure*} 

\begin{figure*}[htbp]
\begin{center}
    \centerline{\includegraphics[width=1.0\textwidth]{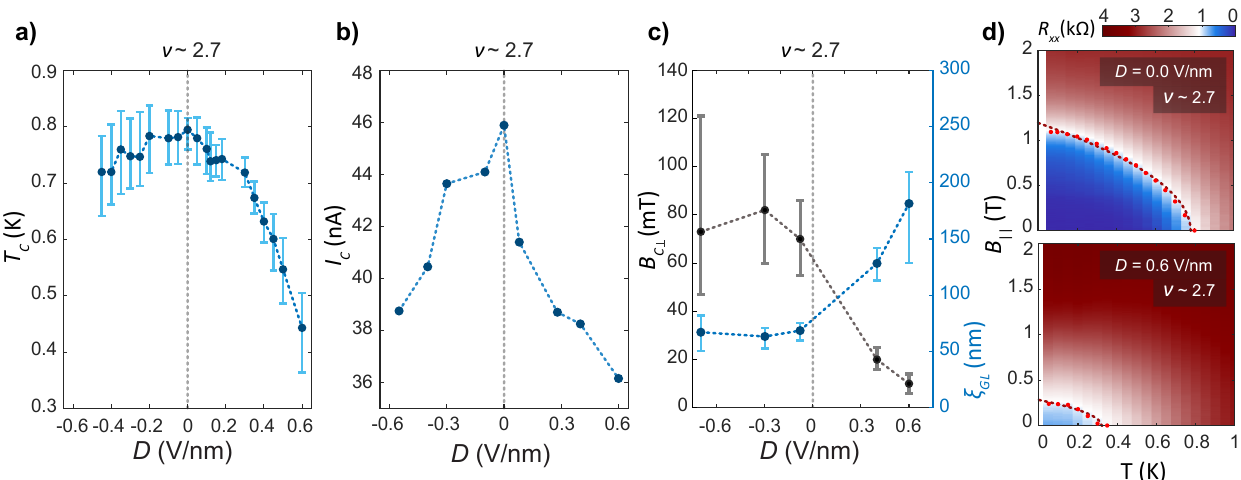}}
\end{center}
\caption{\label{mf4}\textbf{Superconductivity metrics with $D$.} \textbf{a)} Variation of $T_c$ with $D$ at $\nu\sim+2.7$. The values of $T_c$ were approximated as the temperatures at which  $R_{xx}$ drops to $50\%$ of normal state resistance, $R_n$, for the respective $D$ fields. The error bars represent the temperatures corresponding to $60\%$ and $40\%$ of $R_n$. \textbf{b)} Variation of $I_c$ with $D$ at $\nu\sim+2.7$. \textbf{c)} Variation of $B_{c\perp}$ (left axis) and corresponding Ginzburg-Landau coherence length, $\xi_{GL}$ (right axis), with $D$ at $\nu\sim+2.7$. The error bars for $B_{c\perp}$ represent the required magnetic field, $B_\perp$, to cross $60\%$ and $40\%$ thresholds of $R_n$. \textbf{d)} $R_{xx}$ versus $B_\parallel$ and T for zero $D$ (top panel) and $0.60$ V/nm (bottom panel). 
The temperature dependence of $B_{c \parallel}$ (red dots) is fitted (dashed black line) to extract the critical parallel magnetic field at zero temperature, $B^0_{c \parallel}$ $\sim 1.2$ T and $0.3$ T for $D =0.00$ V/nm and $0.60$ V/nm, respectively, which matches well with the Pauli limit calculated from $B_P =1.76 k_{B} T^0_c \sqrt{2}/g\mu_B $, where $T^0_c$ is the zero magnetic field critical temperature, $g$ is the Land\'e g-factor, and $\mu_B$ is Bohr magneton. 
}
\label{Figure4}
\end{figure*} 

\subsection*{Tuning the strength of the SC phase with $D$}
\begin{spacing}{2}
In this section, we discuss our main results on tuning the SC with $D$. Figure.~\ref{mf3}a shows the variation of $R_{xx}$ with $\nu$ at the base temperature 
for several $D$. The corresponding $2-$d colormap of $R_{xx}$ with $\nu$ and $D$ is shown in \textcolor{purple}{SI-Fig. 2a}. As shown in Figure.~\ref{mf3}a, with increasing positive $D$, the resistance peak values at the CNP and full band filling ($\nu = \pm 4$) slightly decrease while the most dramatic effect with $D$ can be seen on the electron-doped side, particularly between $\nu \sim 1-3$. The superconducting region's width around $\nu \sim +2.7$ gradually shrinks as the $D$ value crosses $\geq +0.25$ V/nm. As SC weakens with $D$, a resistance peak emerges at $\nu \sim +2$ together with an increase in the background resistance value around $\nu \sim +1$. This could be an indication of the onset of correlation physics, which is discussed in the next section while presenting the Hall density measurements. The concomitance between the SC suppression and resistance peak appearance with increasing $D$ suggests that the two phases compete rather than being intimately connected. Note that the strength of SC also decreases with negative $D$, but the effect remains weaker compared to the positive $D$ as discussed in \textcolor{purple}{SI- 3} (see \textcolor{purple}{SI-Fig. 2b}). The superconducting region being tuned by $D$ has already been reported in other systems like alternatingly twisted 
magic-angle trilayer 
graphene, 
~\cite{hao2021electric,park2021tunable} quadra layer 
and pentalayer 
~\cite{park2022robust,zhang2022promotion} graphene, but not so far seen in tBLG. Ref~\cite{yankowitz2019tuning} on MAtBLG reports an appearance of a resistive peak at half filling on the hole side ($\nu=-2$) with increasing $D$ (negative to positive). The apparent asymmetric behaviour in the SC phase with $D$ (near $|\nu|\sim2$) is attributed to the structural inhomogeneity in the device but there are no indications of the nature of modification of the strength of the superconducting phase with $D$ in the study~\cite{yankowitz2019tuning}. Here, we have tried to uncover the nature of the tuning of the strength of the SC phase with $D$ in the twisted bilayer graphene below magic-angle.

The tunability of the SC phase in our near magic-angle tBLG can be further seen in the superconducting dome, as shown in Figure.~\ref{mf3}b and ~\ref{mf3}c. It can be seen that compared to Figure.~\ref{mf2}b at $D = 0.00$ V/nm, at $D =0.40$ V/nm and  $0.60$ V/nm in Figure.~\ref{mf3}b,c both the width (in $\nu$) of the SC region and $T_{c}$ are reduced significantly. Figure.~\ref{mf3}d shows $R_{xx}$ versus $T$ at the optimal doping ($\nu \sim +2.7$) for several $D$, emphasizing the tunability of SC with $D$. In Figure.~\ref{mf4}, we have summarized the tunability of several SC metrics with $D$. The variation of $T_c$, $I_c$, and $B_{c\perp}$ with $D$ are shown in Figure.s ~\ref{mf4}a, ~\ref{mf4}b, ~\ref{mf4}c respectively. Raw data related to these plots are shown in \textcolor{purple}{SI- 5} and \textcolor{purple}{SI- 6} (\textcolor{purple}{SI-Fig. 4-9}); which also describes the method ($R_{xx}$ value to one-half of its normal state value\cite{lu2019superconductors,su2023superconductivity,park2021tunable}) used to extract $T_c$, $I_c$, and $B_{c\perp}$. The right side y-axis of Figure. \ref{mf4}c shows the variation of the superconducting coherence length ($\xi_{GL}$) with $D$. At a given $D$, the Ginzburg Landau coherence length ($\xi_{GL}$) is extracted from the inferred zero temperature limit of the critical magnetic field, $B^{0}_{c\perp}$(see \textcolor{purple}{SI- 6 (IV)} for details). Figure.~\ref{mf4}d shows the effect on the SC phase of a parallel magnetic field, $B_\parallel$, and $T$ at $D = 0.00$ V/nm (top panel) and $D = 0.60$ V/nm (bottom panel). The temperature dependence of $B_{c \parallel}$ (red dots in Figure.~\ref{mf4}d) is fitted to the phenomenological relation: $T/T^0_c = 1 - (B_{c\parallel} / B^0_{c\parallel})^2$, 
where $T^0_c$ is the superconducting critical temperature at zero $B$, and the fitting gives the critical parallel magnetic field in the zero temperature limit, $B^0_{c \parallel}$, which matches well with the Pauli 
limit, $B_P$,\cite{su2023superconductivity,cao2021pauli,park2021tunable} 
(see \textcolor{purple}{SI- 6 (V)} for details) at $D = 0.00$ and $0.60$ V/nm. 
Note that the asymmetry in SC metrics with positive and negative $D$ seen in our experiment in Figure.~\ref{mf4} could be related to the asymmetry in the dielectric environment of the top and bottom layer graphene (different thicknesses of the top and bottom hBN, metallic top gate versus SiO$_2/$Si back gate, and 
discussed in detail in \textcolor{purple}{SI- 2} and \textcolor{purple}{SI- 3 (I)}), and has been reported for alternatingly twisted trilayer graphene~\cite{hao2021electric,park2021tunable}.
\end{spacing}

\begin{figure*}[htbp]
\begin{center}
    \centerline{\includegraphics[width=1.0\textwidth]{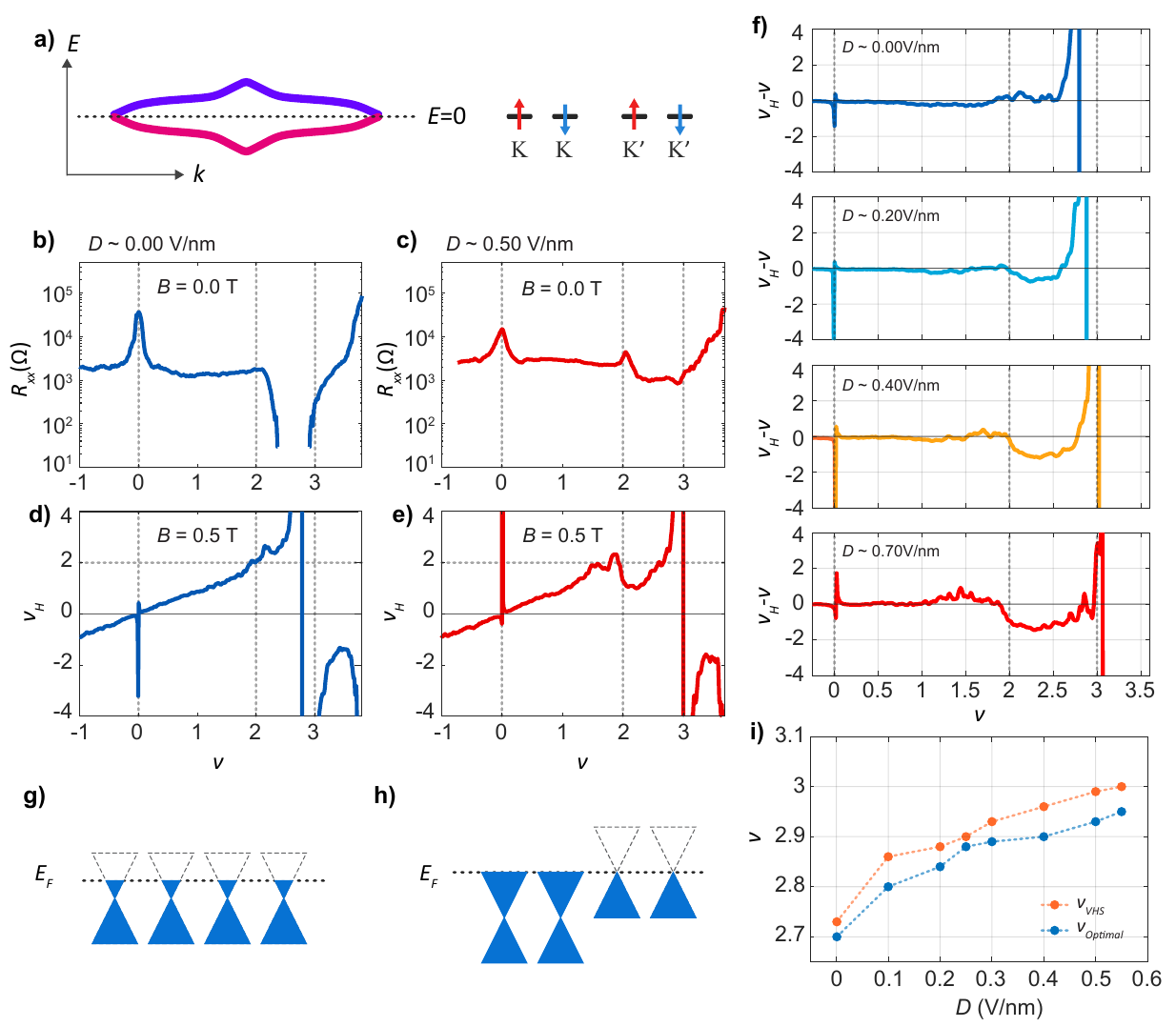}}
\end{center}
\caption{\label{mf5}\textbf{Hall density and isospin symmetry breaking with $D$.} \textbf{a)} Schematic representation of isospin unpolarized single particle low energy degenerate bands in tBLG. \textbf{b)},  \textbf{c)} Longitudinal resistance, $R_{xx}$ vs the filling factor, $\nu$, for displacement field, $D = 0.00$ V/nm and $0.50$ V/nm respectively. \textbf{d)}, \textbf{e)} Normalized Hall density $\nu_H$ 
vs $\nu$ for the same $D$ values as in \textbf{(b)}, \textbf{(c)}. 
\textbf{f)} $\nu_H-\nu$ vs $\nu$ plots for several $D$. 
\textbf{g)} Schematic representation of unpolarized isospin flavors at zero $D$ when all the Dirac-like bands are equally populated 
with degeneracy $g_d = 4$.  \textbf{h)} Schematic representation for polarized isospin flavors with degeneracy $g_d = 2$. 
\textbf{i)} Filling factors of the optimal doping point of the superconducting phase, $\nu_{optimal}$ and the van Hove singularity (vHs) position, $\nu_{vHs}$ with $D$. Both $\nu_{optimal}$ and $\nu_{vHs}$ track each other and increase with increasing $D$.}
\label{Figure5}
\end{figure*} 

\subsection*{Hall density and isospin symmetry breaking with $D$} 
\begin{spacing}{2}
So far, we have discussed how the SC is tuned with $D$; however, to see its connection with the characteristics of the band structure, in this section, we present the Fermiology measurements like the Hall density ($n_H$). 
The single particle low-energy band of tBLG has four flavors, corresponding to two valleys and two spins, as schematically shown in Figure.~\ref{mf5}a. Figure.~\ref{mf5}d shows the variation of the measured Hall filling, $\nu_H$ ($\equiv 4n_H$/$n_s$) (see \textcolor{purple}{SI- 7 (II)} for details), with the moir\'e filling factor, $\nu$, at $D = 0.00$ V/nm on the electron-doped side at $B_\perp = 0.5$ T (anti-symmetrized by measuring at $B_\perp = \pm 0.5$ T, see \textcolor{purple}{SI- 7} for details). At $B_\perp = 0.5$ T, the SC is killed, and it can be seen from Figure.~\ref{mf5}b that $\nu_H$ increases with a linear slope until $\nu \sim +2$, and beyond that, $\nu_H$ increases rapidly and diverges (with value   $>\pm200$, see \textcolor{purple}{SI-Fig. 11a}) on either side of $\nu \sim +2.7$, where it changes sign. This is a distinct clear signature of (the Fermi level crossing) a vHs\cite{hao2021electric}. 
It can be seen from the $R_{xx}$ versus $\nu$ plot in Figure.~\ref{mf5}b ($B_\perp = 0$ T) that the SC phase arises around the vHs, and the linear slope in the Hall filling (versus filling) measurement confirms that all the four flavours are equally populated with the charge carriers as shown schematically in Figure.~\ref{mf5}g, implying that the SC arises from an isospin-unpolarized band. A similar result for twisted double bilayer graphene stabilized by WSe$_2$ \cite{su2023superconductivity} has recently been reported, where the superconductivity emerges from unpolarized states near the van Hove singularities (vHs). Figure.s~\ref{mf5}c,e show similar plots for $D = 0.50$ V/nm, in which case, initially, the $\nu_H$ increases with slope one; however, before reaching the vHs at $\nu\sim +3.0$, there is a suppression of $\nu_H$ around $\nu \sim +2$, where a resistance peak ($R_{xx}$) concomitantly emerges as shown in Figure.~\ref{mf5}c, with no SC. The $2-$d colormap of $\nu_H$ with $\nu$ and $D$ is shown in \textcolor{purple}{SI-Fig. 12a}. To understand the suppression in Hall filling with $\nu$, in Figure.~\ref{mf5}f we plot  $\nu_H - \nu$ versus $\nu$, which remains close to zero at $D = 0.00$ V/nm, reconfirming the degeneracy of $g_d = 4$ of the occupied bands. As can be seen from the figure, with increasing $D$ the values of $\nu_H - \nu$ deviate from zero and become negative between $\nu \sim +2$ and vHs, and their values head towards $\sim -2$, attaining the value of $\sim -1.6$ at $D \sim 0.70$ V/nm, which indicates broken isospin symmetry with $g_d \sim 2$ as schematically shown in Figure.~\ref{mf5}h. 
\end{spacing}

\section*{Discussions}
\begin{spacing}{2}
To summarize the above section, for our device, SC arises around the vHs from an unpolarized band but becomes weaker with broken isospin symmetry (or cascaded transition $/$ Dirac revival \cite{zondiner2020cascade,wong2020cascade}) at $\nu\sim+2$, which is responsible for the appearance of the peak in $R_{xx}$. Though we do not see any signatures of an insulating behaviour of the peak emerging with increasing $|D|$ at $\nu\sim+2$, these two competing phases, SC and the phase with isospin polarization are stabilized respectively for small and large values of the displacement field. A recent theoretical study reported in Fig. S$4$a of ref\cite{long2024evolution} also supports the decrement of superconducting $T_c$ with electric field for a symmetry-broken state ($g_d\sim2$) compared to the completely degenerate ground state with $g_d\sim4$ at zero $D$ $(\equiv$ zero $E)$. Though with increasing $D$ the SC phase gets suppressed, the intimate connection between the diverging density of states (vHs) and SC can be 
seen in Figure.~\ref{mf5}i, where the filling values corresponding to the vHs ($\nu_{vHs}$) and optimal doping for the SC ($\nu_{optimal}$) are plotted versus $D$ and show a close correspondence. In our theory, using a tight-binding model (see \textcolor{purple}{SI- 13}), we capture the shift of the vHs with increasing $D$ as shown in \textcolor{purple}{SI-Fig. 23}. Although the calculated position of fillings corresponding to the vHs do not precisely match experimental observations, the theory captures the essential features qualitatively. Effect of the shift of the vHs to higher Fermi level, $\mu$ with increasing electric field, $E$, and broadening of the density of states (DOS) around it are calculated in ref\cite{long2024evolution} (Fig. S4a) using the continuum model \cite{bistritzer2011moire,lopes2007graphene,koshino2020effective} for a MAtBLG ($\theta=1.1^0$). 
Broadening of the low energy bandwidth and shift of the vHs with the applied displacement field, $D$, in experiments, has also been reported in other twisted platforms such as in twisted bilayer WSe$_2$ \cite{wang2020correlated}.
We want to mention here that there are several other systems, such as rhombohedral trilayer graphene\cite{zhou2021superconductivity}, alternatingly twisted trilayer\cite{hao2021electric,park2021tunable} and multilayer graphene\cite{park2022robust,zhang2022promotion}, in which the SC phase emerges directly from a phase where only two out of the four isospin components are occupied. In comparison, in twisted bilayer graphene, there have been contrasting results between MAtBLG, and detuned \cite{saito2020independent} or screened tBLG\cite{stepanov2020untying,liu2021tuning}. In this work, in a tBLG slightly away from the magic angle, we have shown that upon tuning with $D$, the two phases, SC and the symmetry broken phase, seem to compete rather than being intimately connected.

One could, in principle, argue that the modification with $D$ of the SC phase could be due to the formation of superconducting percolation networks ~\cite{yankowitz2019tuning} arising from the possible presence of disorders. Here we would like to point out that we see, in Figure.~\ref{mf5}, a reduction of Hall filling, $\nu_H$, at $\nu\sim+2$ with increasing $D$ measured at $B_\perp=0.5$ T ($B_{c\perp}\approx60 - 70$ mT at $D=0.00$ V/nm) where no superconductivity survives for any magnitude of $|D|$, leading to an isospin polarization with reduced degeneracy. The gradual increase of $R_{xx}$ to a peak and simultaneous reduction of the $\nu_H$ near $\nu\approx+2$ with $D$, the synergy of both these phenomena can only be understood through the underlying modification of the low energy band structure of below magic-angle tBLG and cannot be reconciled with a picture of the percolating network of SC and normal state puddles. Moreover, the presence of a disorder-induced percolation network is expected to show the Fraunhofer pattern, which is not seen in our experiment for the strongest superconducting dome around $\nu\sim+2.7$ at $D=0.00$ V/nm (Figure. \ref{mf2}d - left panel). The Fraunhofer pattern only arises after weakening the superconductivity at higher $D$ as shown for $\nu\sim+2.3$ at $D=0.53$ V/nm (Figure. \ref{mf2}d - right panel). Furthermore, the different thermal cycles (\textcolor{purple}{SI-Fig. 5}) produce the same results in weakening the SC and the reduction in $\nu_H$ at $\nu\sim+2$ with increasing $D$, even though the disorder configurations are expected to be different in different thermal cycles. Further, the possible effect of contacts in our measured $R_{xx}$ is discussed in \textcolor{purple}{SI- 3 (II)} and ruled out its role in weakening SC. These observations indicate that the formation of superconducting percolation networks ~\cite{yankowitz2019tuning} arising from the possible presence of disorders or any other extrinsic origins are not responsible for the observed phenomena, rather our data suggest an intrinsic effect of the low energy band structure modification with electric field as seen in our theoretical calculations and consistent with ref\cite{long2024evolution}.

In summary, our findings emphasize the influence of band structure tuning through an external electric field on the emergence of superconductivity and other orders in near magic-angle twisted bilayer graphene. Our results shine a light on the close connection between the divergent density of states in the electronic structures (i.e. vHs) and the presence of superconductivity in tBLG near magic-angle. Furthermore, we observe that superconductivity diminishes with alterations to the electronic structure when isospin gets polarized. 
Our results are consistent with the recent theoretical study showing the suppression of SC with electric field in tBLG, and highlight how the origin of SC in tBLG is special compared to other twisted and untwisted graphene-based heterostructures\cite{long2024evolution}. 
\end{spacing}

\clearpage
\section*{Methods}

\subsection{Device fabrication and measurement scheme:}
\begin{spacing}{2}
The device consists of hBN-encapsulated twisted bilayer graphene (tBLG) on a SiO$_2/$Si substrate. The typical length and width of the devices are $ 6$ $\mu$m and $ 2$ $\mu$m, respectively. The usual `tear and stack' technique~\cite{cao2018correlated1,cao2018unconventional1} with a modification is used to fabricate the device and is described in detail in the \textcolor{purple}{Supplementary Information} (\textcolor{purple}{SI- 1}). We measured low-temperature transport at $20\sim25$ mK in a cryo-free dilution refrigerator. Using a constant bias current of $I\leq5$ nA,
the four-probe longitudinal ($V_{xx}$) and transverse voltage ($V_{xy}$) were measured (Figure. \ref{mf1}e) using a lock-in amplifier at a low frequency ($\sim13$ Hz).  
\end{spacing}

\subsection{Theory:}
\begin{spacing}{2}
    
A tBLG structure of twist angle $\theta = 0.95^\circ$ was generated using the TWISTER code \cite{naik2022twister}.
The atomic positions were then relaxed in LAMMPS \cite{thompson2022lammps} up to a force tolerance of $10^{-6}$ ev/\AA. 
Classical force fields were used to model the interatomic interactions, with Tersoff \cite{kinaci2012thermal} as the intra-layer potential and DRIP \cite{wen2018dihedral} as the inter-layer potential.
The electronic band structures were computed using a tight binding model with the transfer integrals approximated using the Slater-Koster formalism \cite{moon2012energy}. 
The displacement fields were integrated into the calculations as on-site terms within each layer, enabling the assessment of their impact on the electronic properties of the system. 
Although the doping levels, at which the van Hove singularities (vHs) occurred in our calculations are lower than those observed experimentally, the qualitative behaviour and the evolution of the vHs on the electron side with displacement field are consistent with the experimental findings. Further details are available in the \textcolor{purple} {Supplementary Information} (\textcolor{purple}{SI- 14}).
\end{spacing}

\section*{Associated Content}
\textcolor{purple}{Supplementary Information} is available for this paper.

\section*{Data availability}
\begin{spacing}{2}
    
All the relevant non-analytical line-plot data generated or measured during this study are included in this published article (and its \textcolor{purple}{Supplementary Information} files). Additional information related to this work is available from the corresponding author upon reasonable request.
\end{spacing}

\section*{Code availability}
\begin{spacing}{2}
The code that supports the findings of this study is available from the corresponding author upon reasonable request.
\end{spacing}

\textbf{Author contributions}\\
R.D. contributed to device fabrication. R.D. and A.G. contributed to data acquisition and analysis. A.G. also contributed to developing measuring codes. S.M., H.R.K., S.B., and M.J. contributed to the development of the theory. K.W. and T.T. synthesized the hBN single crystals. A.D. contributed to conceiving the idea and designing the experiment, data interpretation, and analysis. All the authors contributed to the data interpretation and writing of the manuscript.

\textbf{Notes}\\
The authors declare no competing interests.\\

\section*{Acknowledgements}
\begin{spacing}{2}
A.D. thanks Prof. Ashvin Vishwanath, Prof. Mohit Randeria, and Prof. Nandini Trivedi for the useful discussions. R.D. is grateful to Dr. Manabendra Kuiri, Dr. Arup Kumar Paul, Dr. Ravi Kumar, Dr. Saurabh Kumar Srivastav, Souvik Chakraborty, and Ujjal Roy for numerous discussions on the fabrication of twisted heterostructures. R.D. and A.G. thank Ujjal Roy, Dr. Saurabh Kumar Srivastav, Dr. Ravi Kumar,  and Souvik Chakraborty for their assistance while doing measurements in the He$^3$ cryo-stat (Oxford Heliox) and cryo-free dilution refrigerator (Oxford Triton and Oxford Proteox). H.R.K. and M.J. gratefully acknowledge the National Supercomputing Mission of the Department of Science and Technology, India, and the Science and Engineering
Research Board of the Department of Science and
Technology, India, for financial support under Grants
No. DST/NSM/R\&D HPC Applications/2021/23 and
No. SB/DF/005/2017, respectively. H.R.K. also acknowledges support from the Indian National Science Academy under its grant no. INSA/SP/SS/2023/.  A.D. thanks the Department of Science and Technology (DST) and Science and Engineering Research Board (SERB), India, for financial support (SP/SERB-22-0387) and acknowledges the Swarnajayanti Fellowship of the DST/SJF/PSA-03/2018-19. A.D. also thanks CEFIPRA project SP/IFCP-22-0005. Growing
the hBN crystals received support from the Japan Society for the Promotion of Science (KAKENHI grant nos. 19H05790, 20H00354
and 21H05233) to K.W. and T.T.
\end{spacing}
   
\pagebreak


    \bibliography{reference.bib}

    \newpage
\thispagestyle{empty}
\mbox{}


\section*{Supplementary material}

# Supplementary Information:
# Electric field tunable superconductivity with competing orders in twisted bilayer graphene near magic-angle

Ranit Dutta[1] [†], Ayan Ghosh[1] [†], Shinjan Mandal[1], K. Watanabe[2], T. Taniguchi[3], H.R. Krishnamurthy[1], Sumilan Banerjee[1], Manish Jain[1] and Anindya Das[1] [*]

[1]*Department of Physics, Indian Institute of Science, Bangalore, 560012, India*
[2]*Research Center for Functional Materials, National Institute for Materials Science, 1-1 Namiki, Tsukuba 305-0044, Japan.*
[3]*International Center for Material Nanoarchitectonics, National Institute for Materials Science, 1-1 Namiki, Tsukuba 305-0044, Japan*

[†] These authors contributed equally to this work
[*] anindya@iisc.ac.in

## Contents

**SI- 1 : Device fabrication, device optical image, device response, and twist angle determination.**

We have used the well-established "tear and stack"[1–7] method, with some modifications, to fabricate our twisted bilayer graphene (tBLG) device. First, we exfoliate hBN (thickness ~ 25 − 30 nm), while graphene flakes are exfoliated on separate $SiO_2$/ Si substrates. We then use the optical microscope to identify the required flakes for hBN encapsulation and the graphene flakes for making a twisted bilayer. After identification, we pick up an hBN flake using a transparent PDMS-polypropyl carbonate (PPC) stamp. We use a hemispherically shaped PDMS on one end of a glass slide and cover it with a PPC film to make the PDMS/PPC stamp [8,9]. This PDMS/PPC/hBN stamp is then used to pick up the 'pre-cut' graphene layers sequentially to form the twisted bilayers. The two graphene layers come from a single larger monolayer graphene flake which we pre-cut using a sharp optical fiber tip under an optical microscope [9,10], unlike in the "tear and stack" method. The pre-cutting eliminates the strain and folding in the graphene flakes arising due to the tearing process. To introduce the twist angle between the graphene layers, we first pick up one part of the 'pre-cut' graphene with the PDMS-PPC-hBN stamp at $T$ ~ $40°$ C, then rotate the substrate containing the other half of the graphene flake by $\theta = 1.0°$ using a rotation stage (with a precision of $\theta = 0.04°$) by maintaining the same pick-up temperature as the first graphene flake. Finally, we pick up another hBN flake at $T$ ~ $55°$ C to encapsulate the tBLG layers. The entire stack is then released from the PDMS/PPC stamp on a freshly cleaved $SiO_2$/ Si substrate (dimension ~ 1 $cm^2$) at ~ $80°$ C. The substrate is further cleaned in acetone for a few hours to remove the PPC residue deposited while transferring the heterostructure to the substrate. The optical image of the twisted heterostructure can be seen in SI-Fig. 1a.

To make the electrical contacts (1−D edge contacts [11]), the substrate is spin-coated (at 3000 rpm) with two layers of negative E-beam resist - one layer of 495A4 and on top of it another layer of 950A4. However, instead of the usual temperature of $180°$ C, each PMMA layer is baked at $120°$ C for 15 minutes. This low-temperature baking is done to avoid thermal relaxation of the twist angle between the two graphene layers. Then the contacts are defined using standard E-beam lithography (EBL). After developing the exposed area of the e-beam resist, the contacts are etched with $CHF_3$−$O_2$ (gas flow rate ratio ~ 10 : 1) plasma [12] followed by thermal deposition (base pressure ~ 1e−7 mbar) of Cr/ Pd/ Au (5 nm/ 10 nm/ 70 nm). For the dual-gated structure, after fabricating the 1−D edge contacts we define the area of the top gate by EBL after another round of spin-coating with resists, and after developing a metallic top gate layer is formed by depositing Cr/ Au (5 nm/ 70 nm) in the defined region on top of the top hBN. We make our device in the shape of standard Hall-bar geometry. We define the shape by EBL and after developing the PMMA exposed by the e-beam, we perform RIE-F (dry etch: fluorine chemistry) to remove the parts of the heterostructure to define the Hall-bar geometry. The optical image of the final device can be seen in SI-Fig. 1b.

**Twist angle of the tBLG:** The twist-angle, $\theta$ of the tBLG device is determined from the carrier number densities at the full band filling peaks with respect to the CNP in $R_{xx}$. The area of the superlattice unit cell for a twist angle $\theta$ is given by $A \sim \sqrt{3}a^2/2\theta^2$ [1–7,9], where $a$ = 0.246 nm is the lattice constant of the monolayer graphene. At the full band filling, the superlattice carrier density, $n_s$, can be expressed as $n_s = 4/A = 8\theta^2/\sqrt{3}a^2$. Using this expression the twist angle, $\theta$, is determined from the densities at the full band filling peaks (Figure. 2a of the main text and SI-Fig. 1c,d). In our case, for the superlattice, the peak densities are at $n$ ~ $\pm 2.15 \times 10^{12}$ $cm^{-2}$, which gives a twist angle of $\theta$ ~ $0.95°$. The calculated $\theta$ translates to a moiré wavelength/ period of $\lambda_m$ ~ 14.47 nm (~ $a/\theta$).

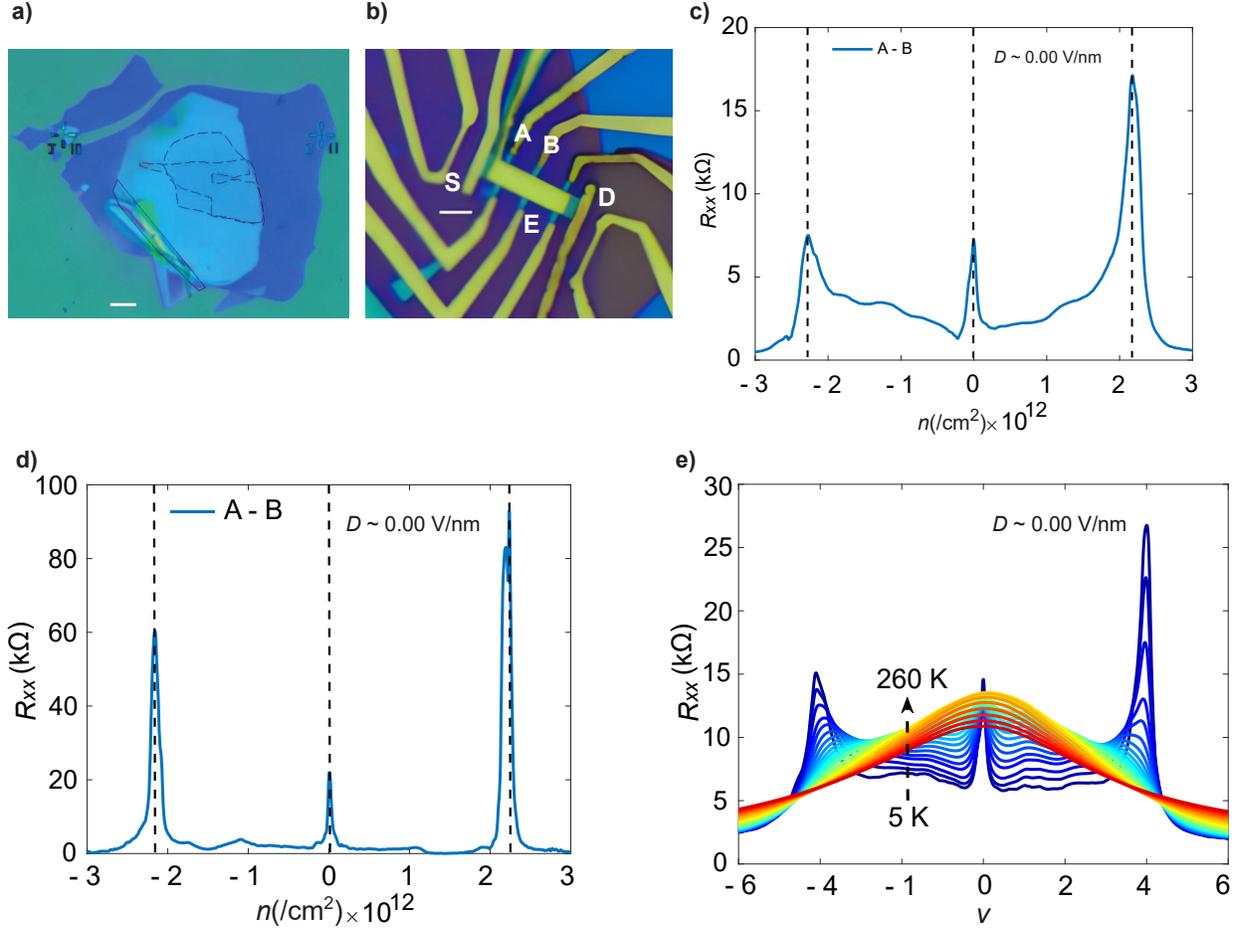

**SI-Fig. 1: Device optical image, response, and twist angle determination:** **a)** Twisted bilayer graphene (tBLG) encapsulated with top and bottom hBN flakes deposited on $SiO_2$/Si substrate. The overlap area of the two twisted graphene layers is marked in dashed lines. The scale in the image is 5 $\mu$m. **b)** Metal contacted device image of a standard Hall-bar with metallic top gate. The scale in the image is 2 $\mu$m. **c)** Four probe resistances ($R_{xx}$ vs $\nu$) at $T$ = 5 K and $D \sim$ 0.00 V/nm between contact probes marked as A and B in **(b)**. Apart from the peak at the charge neutrality point (CNP), two secondary peaks appear almost at the same value of number densities, $\pm n$. **d)** $R_{xx}$ with $\nu$ at 25 mK and at $D \sim$ 0.00 V/nm. The value of the angle extracted from the secondary peaks is $\theta = 0.95° \pm 0.02°$. **e)** $R_{xx}$ vs $\nu$ for various values of increasing temperature from 5 K up to 260 K ($D \sim$ 0.00 V/nm) for contacts A-B measured in the first thermal cycle of the device.

3

## SI- 2 : Controlling the number density, *n*, and the displacement field, *D*.

In this section, we discuss how we have tuned the electrostatic doping and the applied perpendicular displacement field in our measurements. The metallic layer on top of the top hBN and the $SiO_2$/Si act as the top gate and the global back gate, respectively. By sweeping the applied top gate ($V_{tg}$) and back gate ($V_{bg}$) voltages we can control the number density, $n$, and the vertical displacement field, $D$, in our device[13–17].

**Controlling *n*:**
$$n = \frac{C_{bg}(V_{bg} - V_{bg,0}) + C_{tg}(V_{tg} - V_{tg,0})}{e} \quad (1)$$

**Controlling *D*:**
$$D = \frac{C_{bg}(V_{bg} - V_{bg,0}) - C_{tg}(V_{tg} - V_{tg,0})}{2\epsilon_0} \quad (2)$$

Here $C_{bg}$, $C_{tg}$, $V_{bg,0}$, $V_{bg,0}$, $e$ and $\epsilon_0$ are respectively the back gate capacitance per area, top gate capacitance per area, bottom gate voltage offset, top gate voltage offset, bare electronic charge, and free space (vacuum) permittivity. The thicknesses of the encapsulating top and bottom hBN layers are ~ 28 nm and ~ 30 nm, respectively.

## SI- 3 : $\nu - D$ phase diagram from $R_{xx}$.

We have measured the longitudinal resistance, $R_{xx}$, between the contacts A-B by simultaneously changing $V_{tg}$ and $V_{bg}$ by use of a YOKOGAWA GS200 source meter and a Keithley 2400 source meter. The measured $R_{xx}(V_{tg}, V_{bg})$ is converted to a $R_{xx}(\nu, D)$ 2–d colormap using equations (1) and (2) mentioned in SI- 2 and as shown in SI-Fig. 2a. The superconductivity with low resistance on the electronic side (conduction band) and its tunability with $|D|$ are seen.

### I. Discussion on the asymmetric tunability of superconductivity with *D*:

This section discusses why the asymmetry is seen for positive and negative $D$. It can be seen that with $+D$ the superconductivity (observed for the electron doping) is more tunable than for $-D$. This asymmetry may arise from the different dielectric environments surrounding the tBLG. Referring to equation (2) (see SI-Fig. 2b), it is evident that $+D$ directs the electric field from the $SiO_2$/Si substrate towards the metal top gate, while $-D$ directs it from the metal top gate towards the $SiO_2$/Si substrate. As shown schematically in SI-Fig. 2b, the potential energy difference between the layers results in a shift in the Dirac cones of the individual layers. For $+D$, the conduction band of the twisted bilayer graphene (tBLG) originates predominantly from the bottom graphene layer, while for $-D$, the top graphene layer contributes to the conduction band. However, the dielectric environments around the graphene layers are not identical. The bottom graphene layer is separated from the Si layer by the thicker (~ 30 nm) hBN layer and the $SiO_2$ dielectric (≈ 285 nm). In contrast, the top graphene layer is in closer proximity to the metallic top gate, separated only by the thinner hBN layer (~ 27 nm). Due to this asymmetry in the dielectric environment in

our dual-gate structure, the polarizability or screening response of the conduction band would be different for $+D$ and $-D$, which results in the asymmetric response seen in our results.

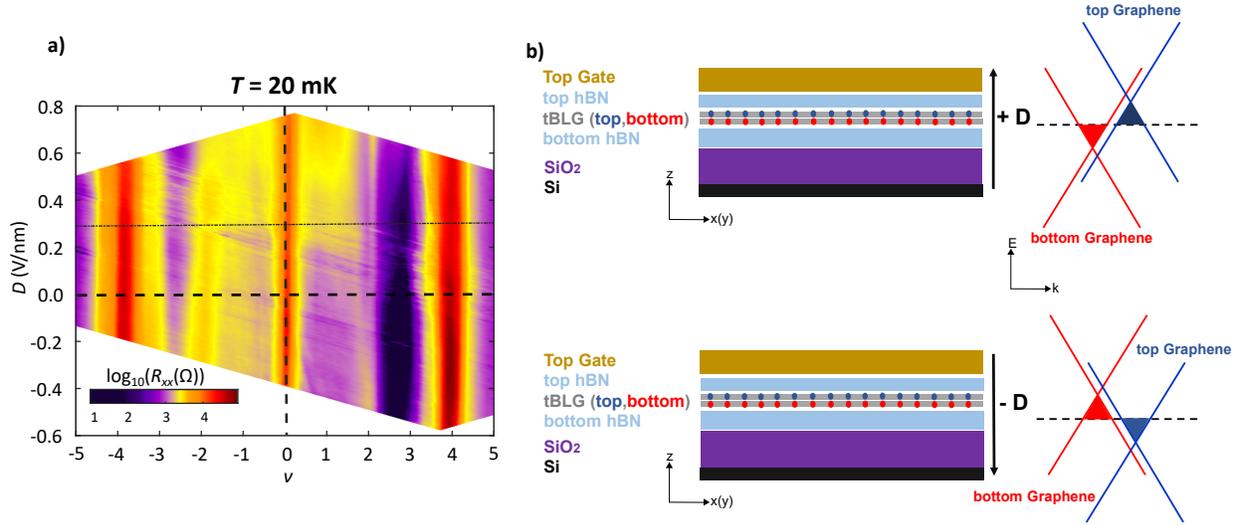

**SI-Fig. 2: $\nu - D$ phase diagram of $R_{xx}$:** **a)** $R_{xx}$ phase map with $\nu$ and $D$ at $T$ = 20 mK. Superconductivity emerges between $\nu = 2$ and $\nu = 3$ on the electron side, as highlighted in dark purple. A slight resistance increase can be seen at $\nu = 2$ as we move towards higher values of $|D|$. **b)** Schematic of the dual-gate structure of the tBLG device. $+D$ points the electric field from $SiO_2/Si$ substrate towards the metal top gate. Similarly, $-D$ points from the metal top gate towards the $SiO_2/Si$ substrate. The right panels show the respective position of the band dispersion coming from the top and bottom layer graphene.

## II. Effect of contacts in the $\nu - D$ phase diagram:

If we look closely into the $\nu - D$ phase diagram in SI-Fig. 2a, between $2 < \nu < 3$ for $+D$ values ranging from ~ 0.15 V/nm to ~ 0.25 V/nm and $D \sim -0.2 V/nm$, the $R_{xx}$ value seems to have a inhomogeneous structure with $D$. This behaviour could easily be understood by considering how the conducting channel, i.e., the current flow through the channel, behaves when only tuned by the Silicone back gate ($V_{bg}$).

SI-Fig. 3a shows the two-probe resistance, $R_{2-probe}$, at 20 mK between device contacts S, D (SI-Fig. 1b and schematic SI-Fig. 3b top panel) with filling factor, $\nu$, tuned only by $V_{bg}$ at $V_{tg} - V_{tg,0}$ = 0 V. SI-Fig. 3c shows the $R_{xx}$ map with $V_{bg} - V_{tg}$ at 20 mK between device contacts A, B (SI-Fig. 1b and schematic SI-Fig. 3b bottom panel). For $\nu \sim 4$, the part of the conducting channel (fabricated out of tBLG) only controlled by $V_{bg}$ becomes highly resistive with $R_{2-probe}$ value reaching more than > 90 kΩ. For $V_{tg} - V_{tg,0}$ = 0 V and the $V_{bg}$ values to tune $\nu$ to ~ 4, by using equation (2), we see that the $D$ values ranges from ~ 0.19 V/nm to ~ 0.22 V/nm (shaded in **pink**). SI-Fig. 3d shows the direction in which the $V_{bg}$, and $V_{tg}$ values are tuned with respect to the $\nu - D$ space.

So, when the arm of the probes (part of the conducting channel not covered by the top gate, hence only tuned by $V_{bg}$) is tuned to such $V_{bg}$ values where $\nu \sim 4$, the arm of the probes measuring $R_{xx}$ become highly resistive, and essentially $V_{xx}$ measurement scheme no longer remain in constant current bias, and thus gives rise to inhomogeneous structure. When we cross $V_{bg} > 35$ V in SI-Fig. 3a (shaded in **brown**) $D$ is > 0.32 V/nm (from equation (2) for $V_{tg} - V_{tg,0}$ = 0 V) and the arm of the probes again become conducting measuring proper $R_{xx}$ resistance of the device. The effect is similar when the arm of the probes are tuned near to $\nu \sim 0$ with $V_{bg}$ with a resistance value exceeding ~ 70 kΩ (marked in ★ in SI-Fig. 3a).

Most of the data relevant to establishing the tuning of the strength of the SC phase with $D$ in our measurement presented in the main text as well as here in the Supplementary Information has been taken and discussed where $R_{xx}$ has a homogenous variation in the $\nu - D$ phase space, for example on $+D$ side we study for $0 < D < 0.10$ V/nm and $D > +0.30$ V/nm and $\nu \sim 2 - 3$. A similar region has been selected for the $-D$ region accordingly for $\nu \sim 2 - 3$ to measure relevant parameters like $T_c$, $I_c$, $B_c$, etc.

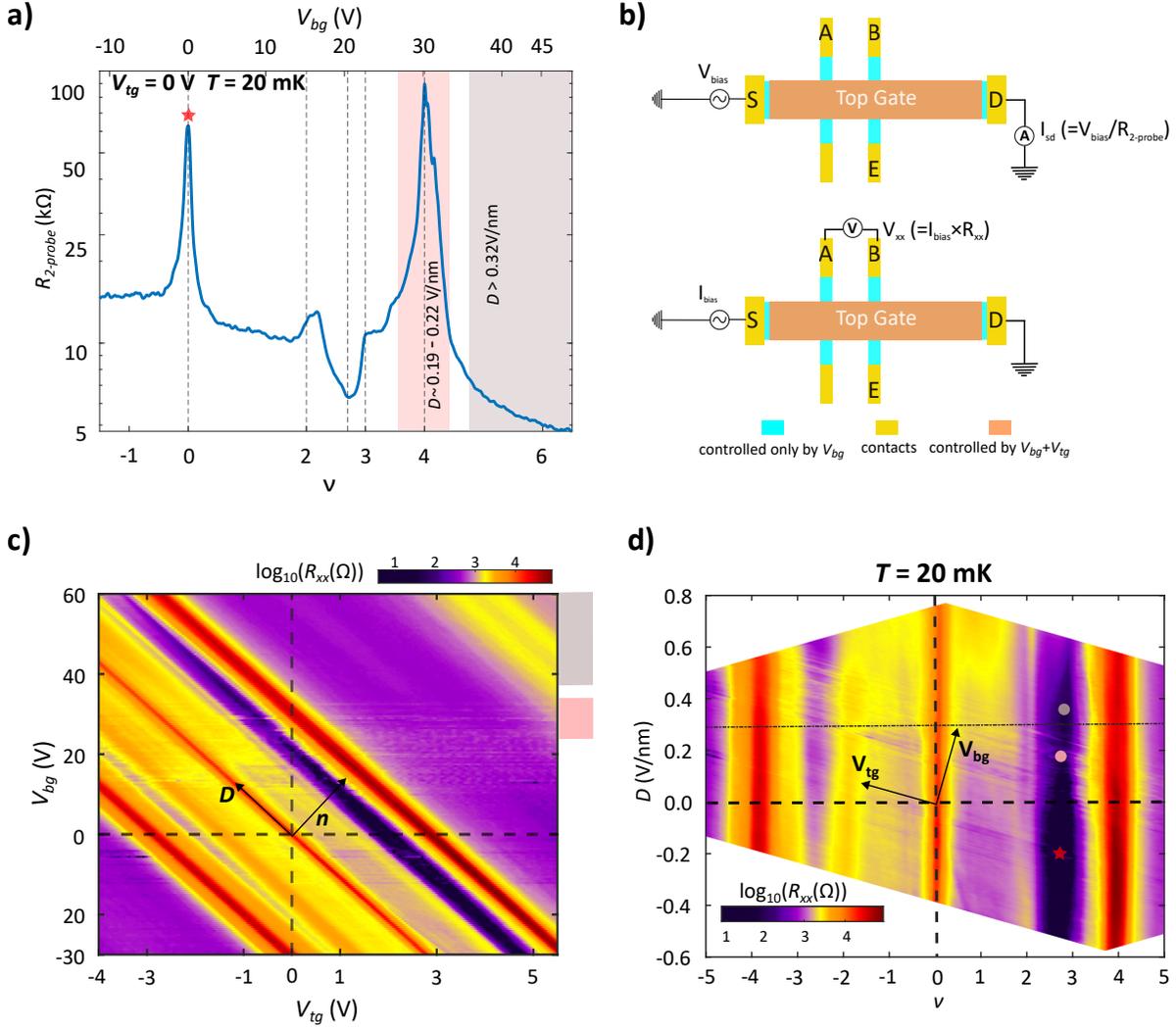

**SI-Fig. 3: Effect of contacts in the $\nu - D$ phase diagram: a)** Two-probe resistance, $R_{2-probe}$, at 20 mK between device contacts 'S', 'D' with filling factor, $\nu$, tuned only by $V_{bg}$ at $V_{tg} - V_{tg,0} = 0$ V. **b)** Schematic representation for measuring $R_{2-probe}$ and $R_{xx}$ of tBLG in our device geometry. **c)** $V_{bg}$–$V_{tg}$ 2-d colormap of $R_{xx}$ measured between 'A' and 'B'. Corresponding $V_{bg}$ regions shaded in **(a)** (pink and brown) are marked on the right vertical axis of the colormap. **d)** Direction of the tuning of $V_{bg}$ and $V_{tg}$ are marked in $\nu - D$ phase space of $R_{xx}$. The position of the corresponding ~ $D$ values of the shaded region in **(a)** are marked with solid circles in respective colours. '★' is the region where the arm of the hall probes are tuned near to $\nu \sim 0$ by $V_{bg}$.

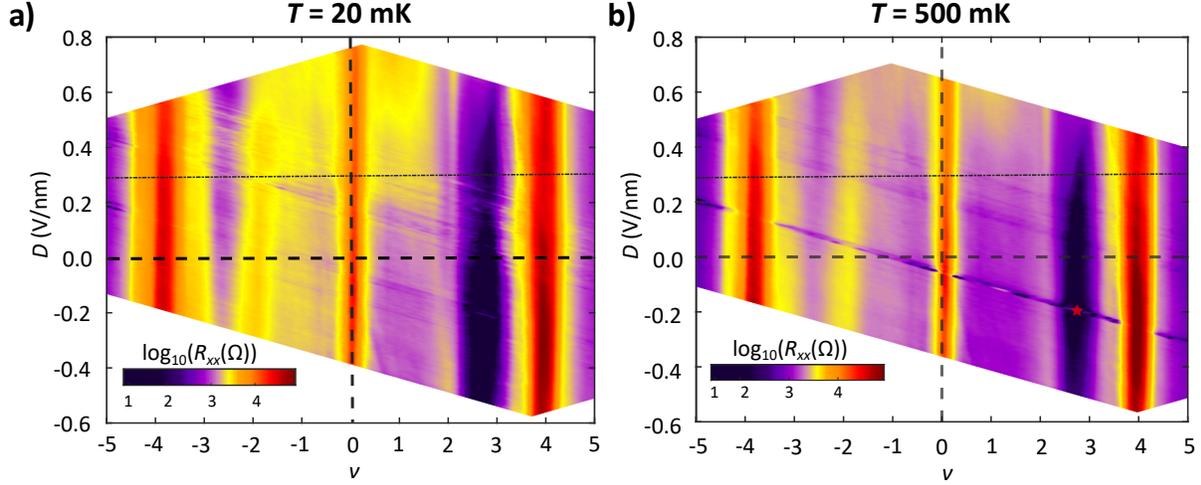

**SI-Fig. 4:** $\nu - D$ **phase diagram in different thermal cycles: a)** Same $R_{xx}$ phase map (at 20 mK) as shown in SI-Fig. 2a. **b)** $R_{xx}$ phase map with $\nu$ and $D$ at $T = 500$ mK measured in a completely different thermal cycle of the device. The diagonal line passing through $2 < \nu < 3$ for $D \sim -0.15$ V/nm to $D \sim -0.20$ V/nm (marked in ★) is due to the arm of the resistance probes are tuned near to $\nu \sim 0$ by $V_{bg}$ making them highly insulating (> 70 kΩ), in a similar fashion as discussed in SI- 3 (II).

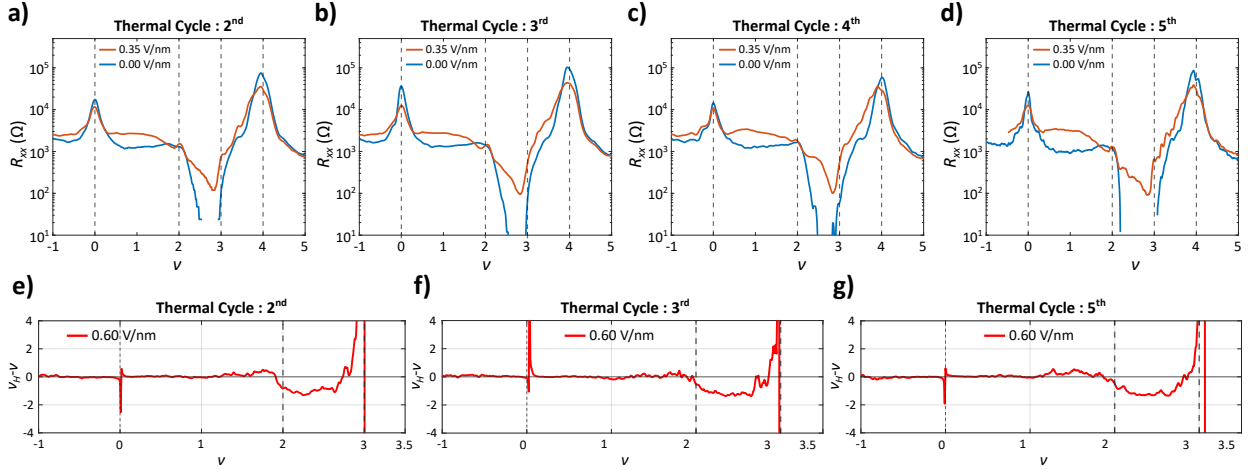

**SI-Fig. 5: Device response in different thermal cycles:** $R_{xx}$ vs $\nu$ measured at $T \sim 20$ mK for $D = 0.00$ V/nm and $0.35$ V/nm in $2^{nd}$ **a)**, $3^{rd}$ **b)**, $4^{th}$ **c)**, and $5^{th}$ **d)** thermal cycle of the device. $\nu_H - \nu$ vs $\nu$ measured at $T \sim 20$ mK for $D = 0.60$ V/nm showing reset of the reduced Hall filling in $2^{nd}$ **e)**, $3^{rd}$ **f)**, and $5^{th}$ **g)**. The sample was warmed to ambient conditions in each thermal cycle before cooling down. $2^{nd}$ and $3^{rd}$ thermal cycles were performed in Oxford Triton dilution fridge. $4^{th}$ and $5^{th}$ thermal cycles were performed in Oxford Proteox dilution fridge.

### III. Observation of SC tuning in repeated thermal cycles:

Similar $R_{xx}(\nu, D)$ was measured at $T = 500$ mK (SI-Fig. 4b) in a different thermal cycle (warming up the device to room temperature and re-cooling to the base temperature). The asymmetric behaviour (of the SC phase) with $D$ in both cases (SI-Fig. 4a and 4b) are similar, which tells us that the tuning of the SC phase is controlled by the displacement field, $D$, in the twisted bilayer graphene below magic-angle.

The device has been subjected to multiple such independent thermal cycles and the observations are consistent in every thermal cycle which is evident from the $R_{xx}$ vs $\nu$ and reduced Hall filling (defined in SI- 7), $\nu_H - \nu$ vs $\nu$ in SI-Fig. 5 at a base temperature of $T \sim 20$ mK. In every such independent thermal cycle, the intrinsic device disorder, if any, will not be configurationally identical throughout the device when it is subjected to a complete 'warm-up' (to ambient room temperature) and subsequent 'full cooldown'. To further elucidate this indifferent effect of disorder on the tunability of the SC phase with $D$, the device was subjected to a couple of independent thermal cycles in completely different cryo-stats as described in SI-Fig. 5.

## SI- 4 : Fraunhofer pattern at different *D*: signature of phase coherence in the SC phase.

2D superconductors show periodic oscillations of the critical bias current, $I_c$, and the differential resistance with varying applied magnetic field, $B_\perp$. This is a result of phase-coherent transport in Josephson junctions formed between the superconducting and insulating regions in the form of an array which is another hallmark of superconductivity in 2D systems [1]. We have measured $dV_{xx}/dI$ with varying $I_{DC}$ with the applied $B_\perp$ ranging from −150 mT to 50 mT at different values of the $\nu$ and $D$ to reveal the signature of phase coherence as demonstrated in Figure. 2d (right panel) in the main text and SI-Fig. 6a,b. The period of oscillations is ∼ 9 − 10 mT.

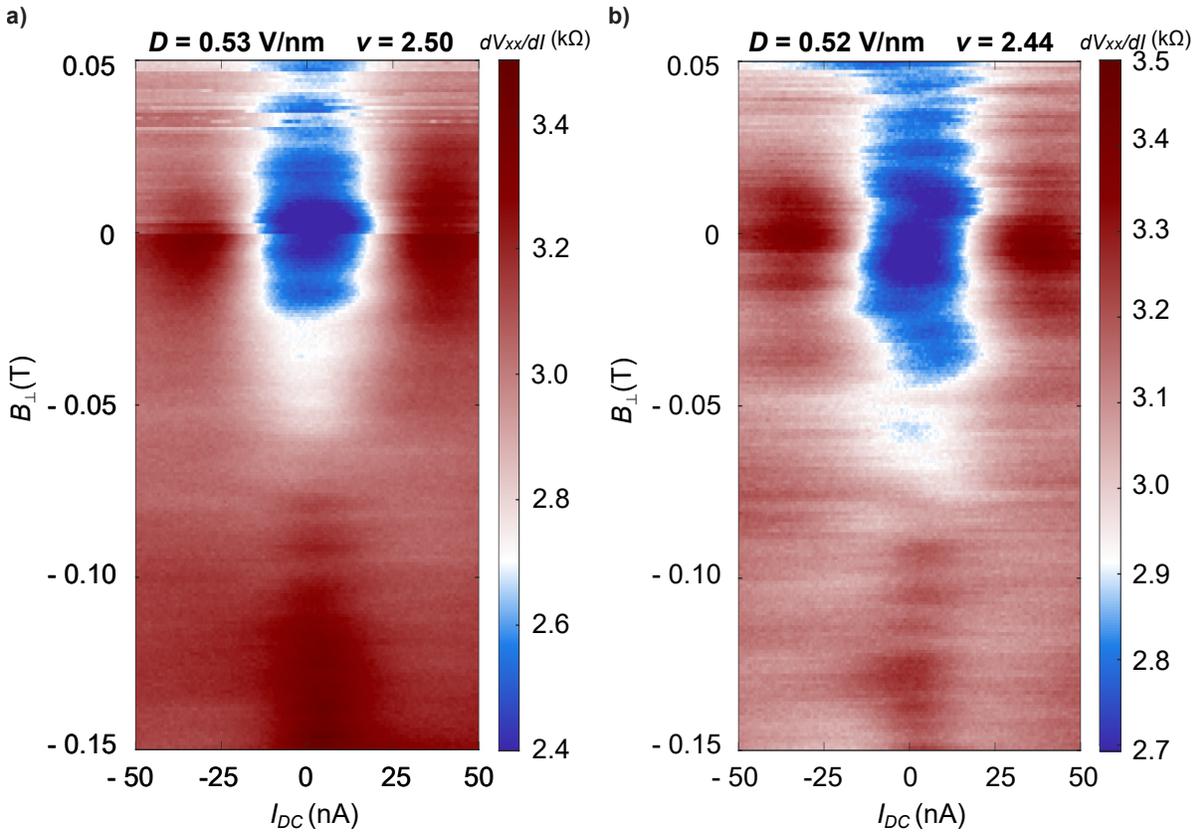

**SI-Fig. 6: Fraunhofer oscillations:** 2−d colormap of the differential resistance with $B_\perp$ and the D.C bias current at $T$ = 25 mK for **a)** $(\nu, D)$ = 2.50, 0.53 V/nm ; and **b)** $(\nu, D)$ = 2.44, 0.52 V/nm . The oscillation period in $\Delta B_\perp$ is ∼ 10 mT and ∼ 9 mT, respectively, which translates to an area of 0.206 $\mu m^2$ and 0.229 $\mu m^2$ respectively for the two chosen $\nu, D$ pairs.

## SI- 5 : Method of extraction of the superconducting critical temperature, T$_c$.

We employed the following method to determine the superconducting critical temperature, $T_c$. SI-Fig. 7a shows $R_{xx}$ vs $T$ plots for $D$ = 0.00 V/nm from the lowest temperature measured in our system of ~ 25 mK up to 1.20 K. We fit the part where $R_{xx}$ starts to saturate ($T \geq 0.9$ K) with a straight line and extrapolate the straight line to $T$ = 0 K. The intercept (defined as $R_{int}$) on the $R_{xx}$ axis at $T$ = 0 K is considered as the resistance of the normal state, $R_n$. We define the superconducting critical temperature, $T_c$, as the temperature $T$ at which $R_{xx}$ drops to 50% of $R_n$.

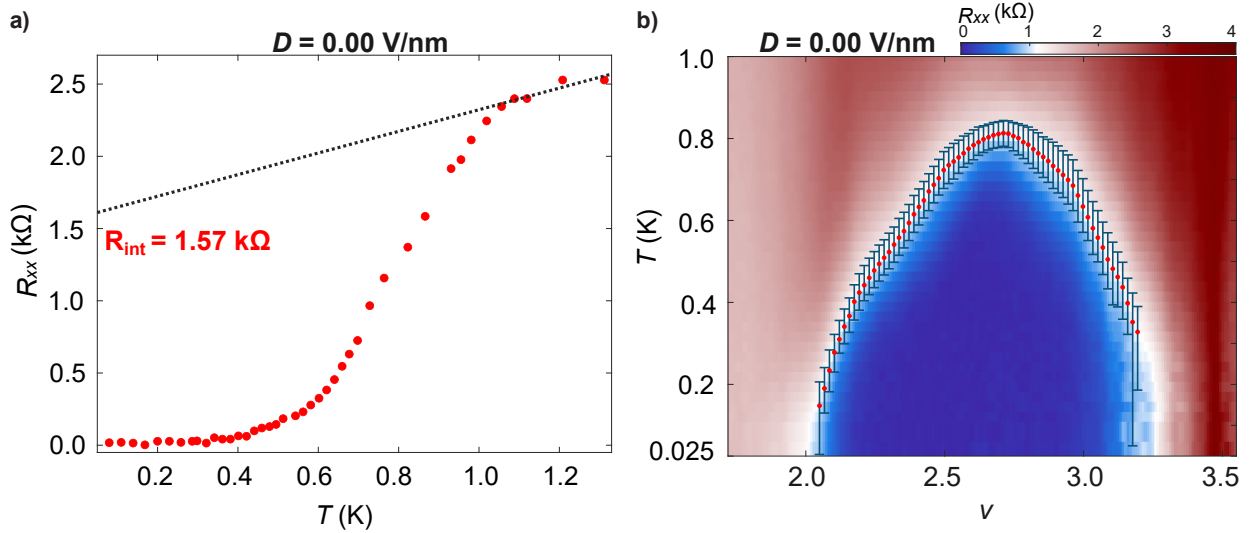

**SI-Fig. 7: Superconducting T$_c$ determination: a)** Increase of the resistance, $R_{xx}$, for $D$ ~ 0.00 V/nm with temperature, $T$, showing the transition from the superconducting to the normal phase at $\nu$ ~ 2.7. The intercept for the straight line that fits the higher temperature ($\geq$ 0.9 K) part is labelled as $R_{int}$ and identified as the normal state resistance, $R_n$. The temperature value for which $R_{xx}$ becomes 50% of $R_n$ is taken as the superconducting critical temperature, $T_c$. **b)** Fixed resistance contour ($\equiv 0.5 R_n$) for $T_c$ projected on the $R_{xx}(\nu, T)$ superconducting dome at $D$ = 0.00 V/nm. The error bars correspond to the 60% and 40% thresholds of $R_n$.

# SI- 6 : Raw data related to superconductivity metrics.

## I. $R_{xx}(\nu, T)$ superconductivity dome at different $D$:

Similar to the SC dome structure shown in SI-Fig. 7b we have measured $R_{xx}(\nu, T)$ domes at other $D$ values as shown in SI-Fig. 8 to see the evolution of the electron SC pocket ($2 < \nu < 3$) at different values of $D$. The variation of $T_c$ seen from these domes is shown in Figure. 4a in the main text.

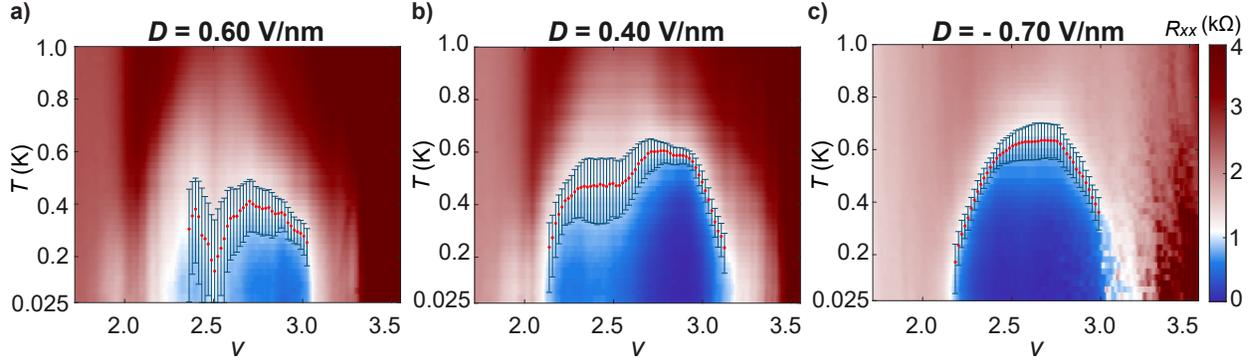

**SI-Fig. 8: $R_{xx}(\nu, T)$ SC dome at different $D$:** 2–d colormap of $R_{xx}$ with $\nu$ and $T$ for the following $D$ values: **a)** 0.60 V/nm, **b)** 0.40 V/nm, and **c)** −0.70 V/nm. Variation of $T_c$ vs $\nu$ can be seen from the projected fixed resistance contour (red dots ≡ $0.5R_n$) at these $D$ fields. The error bars correspond to the 60% and 40% thresholds of $R_n$.

## II. $dV_{xx}/dI$ vs $I_{DC}$ for different values of $D$:

The weakening/suppression of the SC phase with $|D|$ gets reflected in the gradual decrement of the critical DC bias current, $I_c$, as shown in Figure. 4b in the main text. SI-Fig. 9 shows the measured $dV_{xx}/dI$ vs $I_{DC}$ at a few selected $D$ fields. The value of $|I_{DC}|$ where $dV_{xx}/dI$ peaks is taken as $I_c$ for a given $D$.

## III. $R_{xx}(\nu, B_\perp)$ superconductivity dome at different $D$:

$R_{xx}(\nu, B_\perp)$ SC domes at different $D$ values are shown in SI-Fig. 10 to see the evolution of the electron SC pocket at different $D$ as well as to quantify the change in $B_{c\perp}$. The variation of $B_{c\perp}$ as seen from these domes is shown in Figure. 4c in the main text.

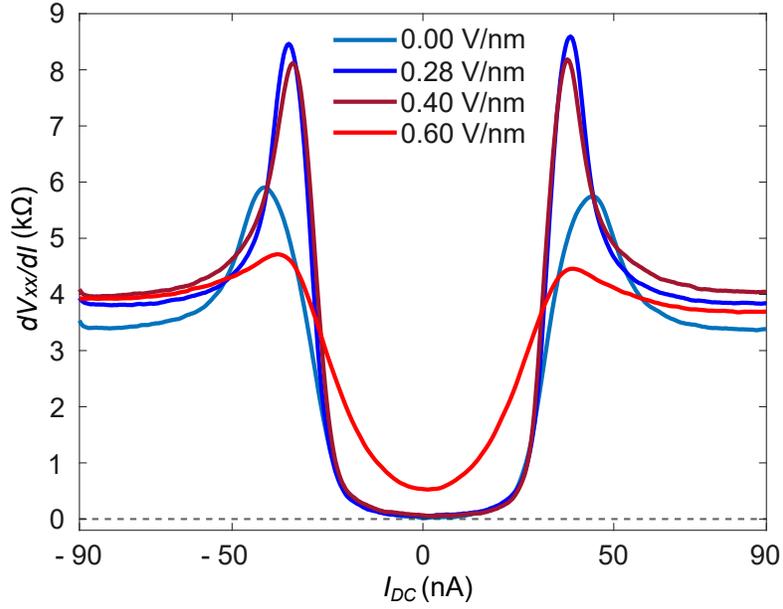

**SI-Fig. 9: Critical D.C. bias current, $I_c$, for different $D$:** Differential resistance is plotted against $I_{DC}$ for different $D$ values. The decrement of $I_c$ values with $D$ is apparent from the shift in the peak positions in $I_{DC}$.

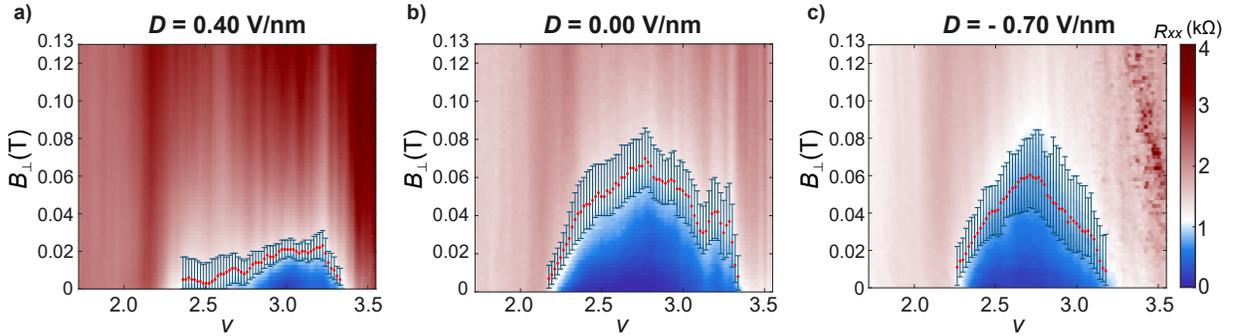

**SI-Fig. 10: $R_{xx}(\nu, B_\perp)$ SC dome at different $\nu - D$:** 2–d colormap of $R_{xx}$ with $\nu$ and $B_\perp$ at 25 mK for the following $D$ values: **a)** 0.40 V/nm, **b)** 0.00 V/nm, and **c)** −0.70 V/nm. Variation of $B_{c\perp}$ can be seen from the projected fixed resistance contour (red dots $\equiv 0.5 R_n$) at these $D$ fields in a similar spirit to the determination of $T_c$ from the $R_{xx}(\nu, T)$ domes in SI-Fig. 8. The error bars correspond to the 60% and 40% thresholds of $R_n$.

13

## IV. Ginzburg-Landau coherence length ($\xi_{GL}$):

From Ginzburg-Landau theory, we can get the dependence of the perpendicular critical magnetic field, $B_{c\perp}$, on temperature which is given by [18] -

$$B_{c\perp} = \frac{\phi_0}{2\pi\xi_{GL}^2}(1 - \frac{T}{T_c}) \tag{3}$$

where $\phi_0 = h/(2e)$ is the superconducting magnetic flux quantum, $h$ is the Planck's constant, $T_c$ is superconducting critical temperature (for a given $D$) and $\xi_{GL}$ is the superconducting coherence length. In SI-Fig. 11, we have shown how $\xi_{GL}$ is extracted from the $R_{xx}(T, B_{c\perp})$ plot. For $\nu, D$ and in the superconducting phase, we measure $R_{xx}$ for different values of $T$ and $B_\perp$ to generate the $T, B_\perp$ phase diagram for a given strength of the SC phase. Using the resistance threshold of $0.5R_n$ as described in SI- 5 we can extract the critical perpendicular magnetic field, $B_{c\perp}$, at each measured $T$. The result is a plot shown in SI- Fig.11b which can be fitted to equation (3) to extract the $T = 0$ K limit of the critical perpendicular magnetic field, $B_{c\perp}^0$ (through the intercept on the $B_{c\perp}$ axis). The GL coherence length for the measured $\nu, D$ is then given as -

$$\xi_{GL} = (\frac{\phi_0}{2\pi B_{c\perp}^0})^{1/2} \tag{4}$$

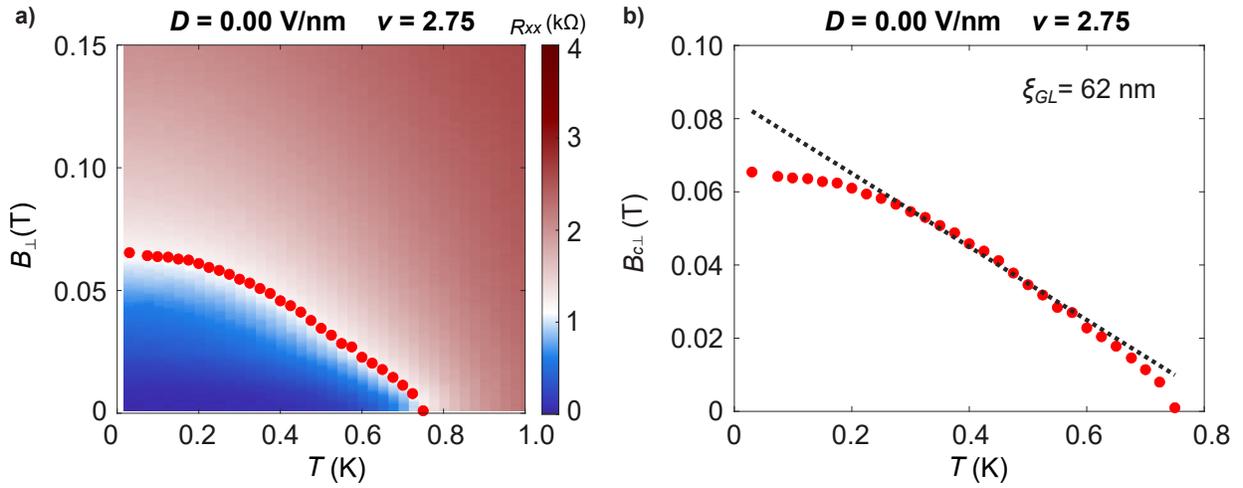

**SI-Fig. 11: $\xi_{GL}$ extraction: a)** $R_{xx}$ colormap of the SC phase at $D = 0.00$ V/nm and $\nu = 2.75$ with $T$ and $B_\perp$. Values of the critical perpendicular magnetic field, $B_{c\perp}$, at each temperature, are marked by red solid circles. **b)** The $B_{c\perp}$ at different $T$ is fitted with a straight line to the Ginzburg - Landau expression (equation (3)) to extract the inferred $T = 0$ K limit of the critical perpendicular magnetic field, $B_{c\perp}^0$. The extracted value of GL coherence length ($\xi_{GL}$) is $\simeq 62$ nm.

14

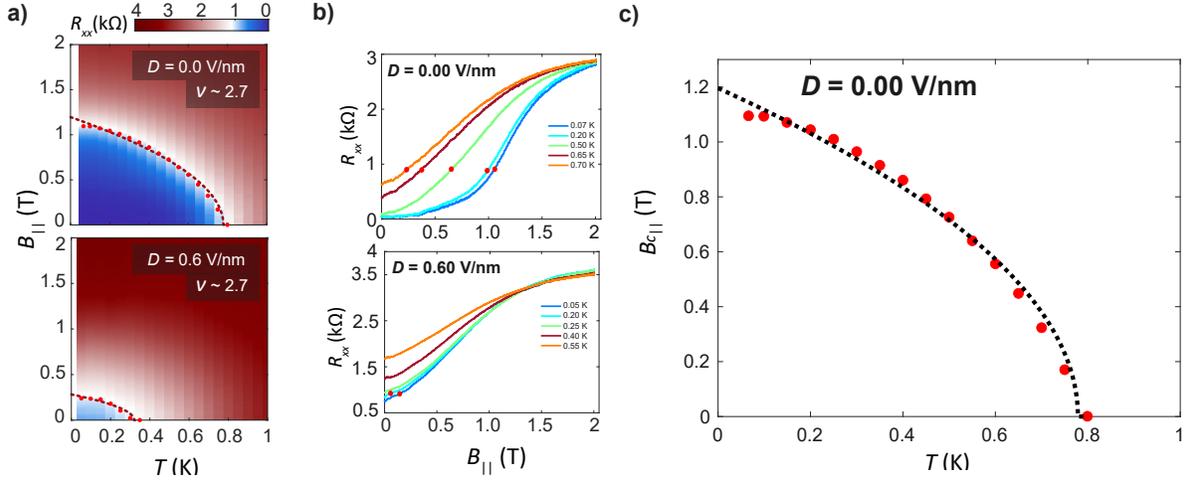

**SI-Fig. 12: $B_{c\parallel}^0$ extraction: a)** $R_{xx}$ with $B_\parallel$ and T for zero $D$ (top panel) and 0.60 V/nm (bottom panel). The temperature dependence of $B_{c\parallel}$ (red circles) is fitted (dashed black line) to extract the critical parallel magnetic field at zero temperature, $B_{c\parallel}^0 \sim 1.2$ T and 0.3 T for $D$ = 0.00 V/nm and 0.60 V/nm, respectively at $\nu \sim 2.7$, which matches well with the Pauli limit calculated from the zero magnetic field $T_c^0$, $B_P = 1.76 k_B T_c^0 \sqrt{2}/g\mu_B$. **b)** Line plots of $R_{xx}$ vs $B_\parallel$ for $D$ = 0.00 V/nm (top panel) and 0.60 V/nm (bottom panel) with increasing $T$. The red circles are the points with resistance values $\simeq 0.5 R_n$. **c)** Fixed resistance ($\equiv 0.5 R_n$) contour points in red circles in $T - B_{c\parallel}$. The dashed black curve is fit to the points by using $T/T_c^0 = 1 - \alpha (B_{c\parallel})^2$, where $T_c^0 \sim 0.8$ K for $D$ = 0.00 V/nm. From the fit $B_{c\parallel}^0 \simeq 1.2$ T for $D$ = 0.00 V/nm.

## V. $B_{c\parallel}$ and Pauli limit ($B_P$): implication, measurement, and calculation.

This section discusses the effect of the in-plane or parallel magnetic field, $B_\parallel$, in a conventional two-dimensional superconductor. In the limit of BCS theory a superconductor with spin-singlet Cooper pairs is split when the Zeeman energy induced by the in-plane magnetic field exceeds the superconducting pairing gap, $\Delta = 1.76 k_B T_c$ [18], where $k_B$ is the Boltzmann constant. The in-plane magnetic field limit for such weakly coupled superconductors is known as the Pauli (or Clogston–Chandrasekhar) limit [19,20] which can be written as $B_P = 1.76 k_B T_c^0 \sqrt{2}/g\mu_B$ [14] and gives $B_P$ = 1.86 Tesla/K $\times T_c^0$ for a Landé g-factor ($g$) of 2. Empirically, the temperature dependence of the in-plane critical magnetic field, $B_{c\parallel}$, is as follows [21]-

$$B_{c\parallel} \propto (1 - \frac{T}{T_c^0})^{1/2} \qquad (5)$$

where $T_c^0$ is superconducting critical temperature at $B_\parallel$ = 0 T (for a given $D$).

For different values of $D$ at $\nu_{optimal}$ in the SC phase, we measure a $R_{xx}$ at a discrete set of temperatures, $T$, while continuously tuning the applied $B_\parallel$ (SI-Fig. 12a). The resistance threshold of $0.5 R_n$ as described in SI- 5 gives the values for the critical in-plane magnetic field, $B_{c\parallel}$ (SI-Fig. 12b), at each $T$ in the form of

15

a fixed resistance contour for $T - B_{c\|}$ (SI-Fig. 12c). The contour at each such $\nu, D$ choice is fitted to $T/T_c^0 = 1 - \alpha(B_{c\|})^2$ [1,3,20,22], where $\alpha$ is a fitting parameter and at the inferred $T = 0$ K limit $1/\sqrt{\alpha} = B_{c\|}^0$, zero temperature critical in-plane magnetic field. We have found that in our measured tBLG device, for a given $D$, $B_{c\|}^0$ is less than the expected value of the Pauli limit, $B_P$ ( for $D$ = 0.00 V/nm $B_P \simeq 1.48$ T and $B_{c\|}^0 \simeq 1.2$ T ), confirming the conventional weakly-coupled nature of the superconductivity in tBLG.

## SI- 7 : $R_{xy}$ and Hall filling, $\nu_H$.

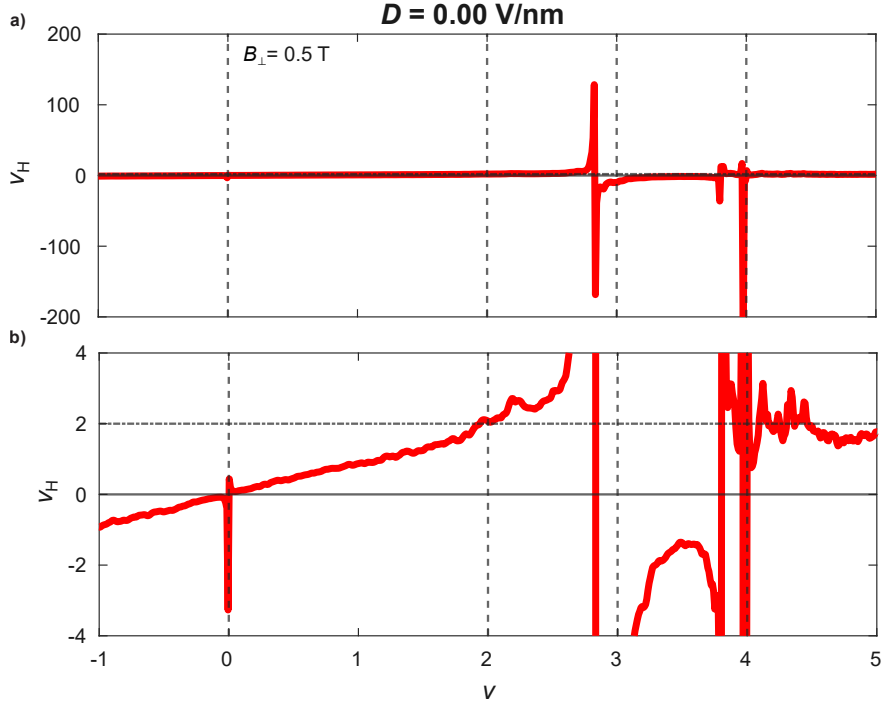

**SI-Fig. 13:** $\nu_H$ **vs** $\nu$: **a)** $\nu_H$ vs $\nu$ at $D$ = 0.00 V/nm. Divergence in $\nu_H$ (position of vHs) is reflected in high $|\nu_H|(\geq 200)$ values. **b)** Zoomed in plot of $\nu_H$ in the range ±4 showing the change in carrier sign across CNP and divergences (vHs) at $\nu \sim 2.7$ and $\nu \sim 4$.

### I. Anti-symmetrization of $R_{xy}$:

The transverse resistance, $R_{xy}$, is anti-symmetrized to avoid the contribution due to the possible asymmetric position of the transverse contacts. The anti-symmetrization is done as follows -

$$R_{xy}^{anti-symm} = R_{xy} = \frac{R_{xy}(+B_\perp) - R_{xy}(-B_\perp)}{2} \quad (6)$$

where, $|B_\perp|$ = 0.5 T.

16

## II. Definition of Hall filling, $\nu_{\text{H}}$:

$$\nu_H = \frac{4n_H}{n_s} = \frac{4|B_\perp|}{eR_{xy}n_s} \tag{7}$$

where, $n_H = |B_\perp|/eR_{xy}$ is the Hall density, $e$ is the bare electronic charge, and $n_s$ is the full band filling number density or the position of the secondary peaks in $R_{xx}$ on the $n$ axis. Whenever $R_{xy}$ becomes zero, $\nu_H$ diverges giving us insight into the van Hove singularities (vHs) in the band structure or the filling landscape. SI-Fig. 13 shows the $\nu_H$ vs the filling, $\nu$, for $D = 0.00$ V/nm (at $|B_\perp| = 0.5$ T). Large divergences in $\nu_H$ are seen near $\nu \sim 2.7$ and $\nu \sim 4.0$. Divergence/vHs at $\nu \sim 2.7$ directly coincides with the position of the SC phase as shown in the main text in Figure. 5b,d.

## SI- 8 : $\nu - D$ phase diagram of normalized Hall density (Hall filling), $\nu_{\text{H}}$.

Similarly to the $\nu - D$ map of $R_{xx}$ mentioned in SI- 3 (SI-Fig.2a), we have measured the transverse resistance, $R_{xy}(+B_\perp)$, and $R_{xy}(-B_\perp)$, between contacts B-E (SI-Fig. 1b) by simultaneously changing the $V_{tg}$ and $V_{bg}$ at $|B_\perp| = 0.5$ T. The anti-symmetrized (equation (6)) $R_{xy}^{anti-symm}(V_{tg}, V_{bg})$ converted to a $\nu_H(\nu, D)$ colormap using equations (1), (2), and (7) is shown in SI-Fig. 14a. SI-Fig. 14b shows a cartoon $\nu, D$ map with all the phases, i.e superconductivity (SC), isospin broken symmetry (ISB), and vHs projected according to the regions in $\nu$ and $D$ where they appear in SI-Fig. 2a and SI-Fig. 14a.

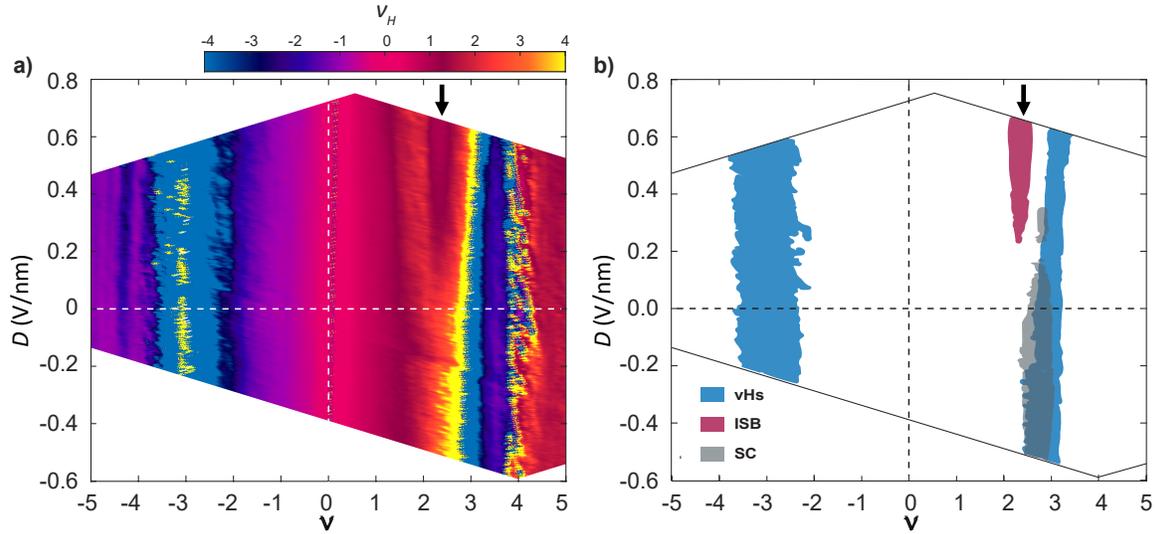

**SI-Fig. 14: Normalized Hall density $\nu_H$ vs $D$: a)** Normalized Hall density or Hall filling, $\nu_H$ ( $\frac{4|B_\perp|}{eR_{xy}n_s}$ ), plotted as a function of $\nu$ and $D$ for $|B_\perp|$ = 0.5 T and $T$ = 25 mK. The vHs appears on the electron side at around $\nu \sim 2.7$ for $D \sim 0.00$ V/nm. The vHs at $\nu \sim 2.7$ migrates towards $\nu \sim 3$ with increasing +ve $D$. For $D \geq +0.20$ V/nm the change in slope in $\nu_H$ vs $\nu$ as seen in Figure. 5f in the main text can be seen from the change in the color gradient around $\nu \sim 2$ marked by a black arrow (↓). The reduction of sudden Hall density (filling) is due to the broken isospin-symmetry (ISB) with increasing $D$. Indication of rapid sign changes with large divergences (vHs) can also be seen on the hole side centered around $\nu \sim -3$ for the whole range of accessible $D$. **b)** Schematic of the $D$ dependent phases with filling factor, $\nu$, observed in our tBLG device. We have used three different colors to mark the overlapping region of superconductivity (gray), vHs (blue), and the regions of Hall density (filling) reduction marked as 'ISB' for the isospin symmetry breaking phase (magenta) also marked by a black arrow (↓) similar to in **(a)**. The 'ISB' phase bounds the two phases (SC and vHs) on the left side of the $\nu - D$ diagram for $D \geq +0.20$ V/nm centred around $\nu \sim 2$.

## SI- 9 : Differential resistance vs $\nu$ at different *D*.

In this section, we discuss another way by which we see the evolution of the electron SC pocket with the application of $D$ in the filling range of $2 \leq \nu \leq 3$. SI-Fig. 15 shows the colormaps of the differential resistance measured as a function of the D.C bias current, $I_{DC}$, and the filling factor, $\nu$, at three different representative displacement fields, −ve $D$ (−0.50 V/nm), neutral $D$ (0.00 V/nm), and +ve $D$ (0.60 V/nm).

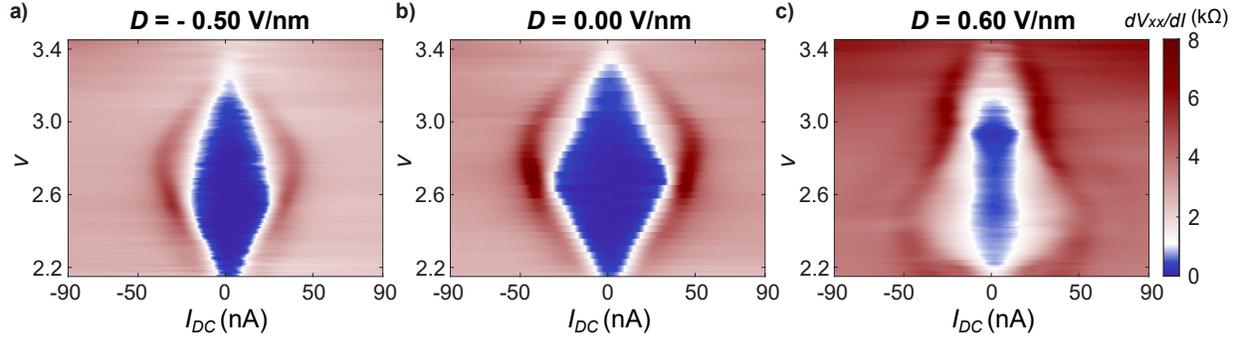

**SI-Fig. 15: $dV_{xx}/dI(\nu, I_{DC})$ at different *D*:** To showcase the variation of superconductivity with +ve and −ve displacement fields, $D$, we have plotted $dV_{xx}/dI$ vs $\nu$ and $I_{DC}$ at the following $D$ values : **a)** −0.50 V/nm, **b)** 0.00 V/nm, and **c)** 0.60 V/nm, with $\nu$ being tuned across the SC region. The shift in the optimal doping point, $\nu_{optimal}$, with the application of $D$ across 0.00 V/nm is evident from the 2−d colormaps.

## SI- 10 : Absence of Anomalous Hall Effect (AHE)

The signature of magnetic structures like anomalous Hall effect [23–25], in tBLG and other twisted systems like twisted double bilayer graphene (TDBLG) [15] has reportedly been observed as hysteresis in transverse resistance, $R_{xy}$, in the presence of the applied magnetic field akin to sharp jumps in resistance (magnetization) at coercive fields. The possibility of the tuning of the superconductivity with $D$ in our device is due to the presence of 'orbital' magnetic structures (magnetism) is ruled out in the hysteresis experiment at zero $D$ and symmetry broken high $D$ regime by sweeping $B_\perp$. We do not see any sign of significant hysteresis in either $R_{xx}$ or $R_{xy}$ (see SI-Fig. 16), barring minimal difference in the magnetoresistance at very small $B_\perp$, confirming the absence of magnetism ordering as well as an anomalous Hall behaviour with $B_\perp$ in our device.

19

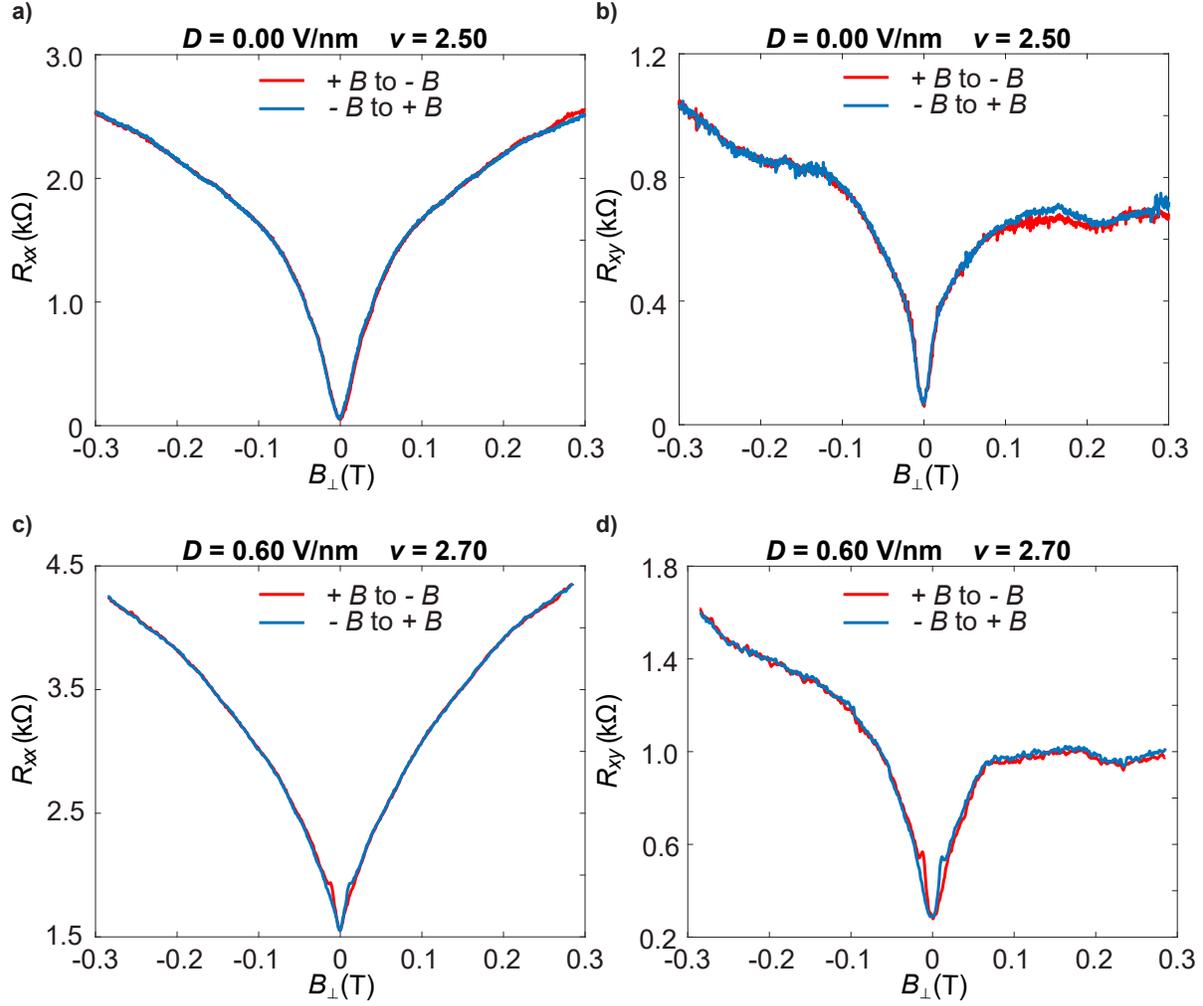

**SI-Fig. 16: Hysteresis measurements with $B_\perp$:** **a)** $R_{xx}$, and **b)** $R_{xy}$ for two sweep directions ($+B_\perp$ to $-B_\perp$) and ($-B_\perp$ to $+B_\perp$) in the range of $B_\perp = \pm 300$ mT (ramp rate ~ 10 mT/min) at $(\nu, D) = 2.5, 0.00$ V/nm. The resistance values are almost similar, with no sign of hysteresis. **c)** $R_{xx}$, and **d)** $R_{xy}$ for $(\nu, D) = 2.7, 0.60$ V/nm also show similar behavior. Here the ramp rate for the magnetic field is ~ 100 mT/min. The small kinks seen in the $R_{xx}$ and $R_{xy}$ magnetoresistance could be due to a finite change in the local temperature of the sample stage due to the higher ramp rate of the superconducting magnet in the cryo-free dilution fridge.

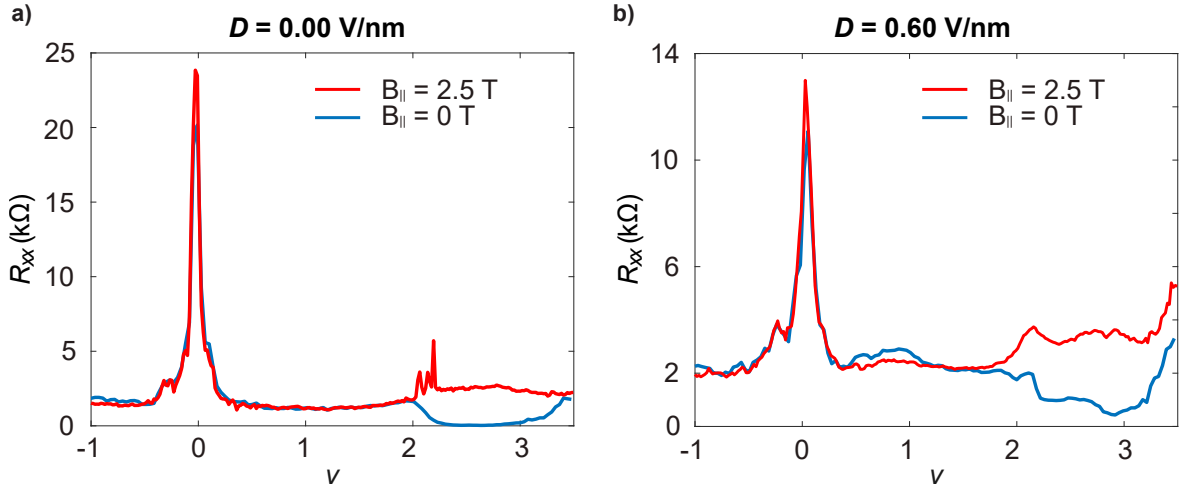

**SI-Fig. 17: $R_{xx}$ in a finite $B_\parallel$:** Appearance of the resistance peak around $\nu \sim 2$ after killing off the SC phase by the application of an in-plane magnetic field exceeding the Pauli limit ($B_P$) for **a)** $D = 0.00$ V/nm, and **b)** 0.60 V/nm.

## SI- 11 : $R_{xx}$ in $B_\parallel$.

In section SI- 6 (V) we have seen how an in-plane magnetic field affects the superconducting phase by splitting the Cooper pairs for magnetic fields above the Pauli limit, $B_P$. $R_{xx}$ vs $\nu$ at two different values of $B_\parallel$ is shown in SI-Fig. 17, where we see the appearance of a resistance peak after the SC phase gets killed by the application of an in-plane magnetic field exceeding the $B_P$ for $D = 0.00$ V/nm and 0.60 V/nm. In our tBLG device, at $B_\parallel = 0$ T there is no signature of correlation peak at zero $D$, though we see a finite resistance increase at $\nu \sim 2$ with the application of $D$ as shown in Figure. 3a in the main text. As discussed in the main text, the weakening of superconductivity with $D$ is likely connected to the flavour polarization of bands due to symmetry breaking near $\nu \sim 2$.

21

## SI- 12 : Theoretical calculations.

### I. Tight binding model:

The electronic band structure is calculated using a tight binding model with the transfer integrals approximated under the Slater-Koster formalism using parameters from Ref. [26,27]. We further take care of the local curvatures at each atomic site introduced by the relaxation effects [28].

$$\hat{H} = -\sum_{i,j} t(\mathbf{r}_{ij}) c_i^\dagger c_j + \text{h.c.}$$
$$t(\mathbf{r}_{ij}) = t_{\pi\pi}[\hat{\mathbf{n}}_i - (\hat{\mathbf{n}}_i \cdot \hat{\mathbf{r}}_{ij})\hat{\mathbf{r}}_{ij}] \cdot [\hat{\mathbf{n}}_j - (\hat{\mathbf{n}}_j \cdot \hat{\mathbf{r}}_{ij})\hat{\mathbf{r}}_{ij}] + t_{\sigma\sigma}[\hat{\mathbf{n}}_i \cdot \hat{\mathbf{r}}_{ij}] \cdot [\hat{\mathbf{n}}_j \cdot \hat{\mathbf{r}}_{ij}]$$
$$t_{\pi\pi} = t_\pi^0 \left(\frac{|\mathbf{r}_{ij}| - a_0}{\delta}\right); \quad t_{\sigma\sigma} = t_\sigma^0 \left(\frac{|\mathbf{r}_{ij}| - z_0}{\delta}\right)$$
$$t_\pi^0 = -2.7\text{eV} \quad t_\sigma^0 = 0.48\text{eV} \quad z_0 = 3.35\text{Å} \quad a_0 = 1.42\text{Å} \quad \delta = 0.184\sqrt{3}a_0 \quad (8)$$

In this context, $\mathbf{r}_i$ represents the real space position of the $i^{\text{th}}$ atom, while $c_i^\dagger$ and $c_i$ stand for the creation and annihilation operators related to the $p_z$ Wannier orbitals at $\mathbf{r_i}$, respectively. The vector $\mathbf{r}_{ij} = (\mathbf{r}_i - \mathbf{r}_j)$ denotes the displacement between the positions of atoms $i$ and $j$, and the unit normal at the $i^{\text{th}}$ site is denoted by $\hat{\mathbf{n}}_i$. Additionally, $\hat{\mathbf{r}}_{ij} = \frac{\mathbf{r}_{ij}}{|\mathbf{r}_{ij}|}$ signifies the unit vector along the direction from atom $j$ to atom $i$.

### II. Incorporating displacement field, $D$:

The effect of the electric field at each atomic site is incorporated via onsite energies

$$\epsilon_i = \begin{cases} -\frac{\Delta}{2} & \text{if } i \in \text{ bottom layer} \\ \frac{\Delta}{2} & \text{if } i \in \text{ top layer} \end{cases} \quad (9)$$

with $\Delta = \frac{D}{\epsilon_{hBN}} z_0$, where $D$ is the value of the perpendicular displacement field in V/nm and $\epsilon_{hBN} \approx 3.8$ is the dielectric constant of hexagonal boron nitride (hBN) (see SI-Fig. 18).

The density of states (SI-Fig. 19) and the number densities are calculated on a $(40 \times 40)$ $\mathbf{k}$ grid through the linear triangulation method, which corresponds to the two-dimensional equivalent of the linear tetrahedron method. The theoretically determined van Hove singularities (vHs) are marked as dashed lines on the electron side. While the computed positions of the vHs deviate slightly from the experimentally observed positions, the qualitative behaviour of the vHs on both the electron and hole sides, in response to the increasing displacement field, $D$, is in excellent agreement with the experimental findings.

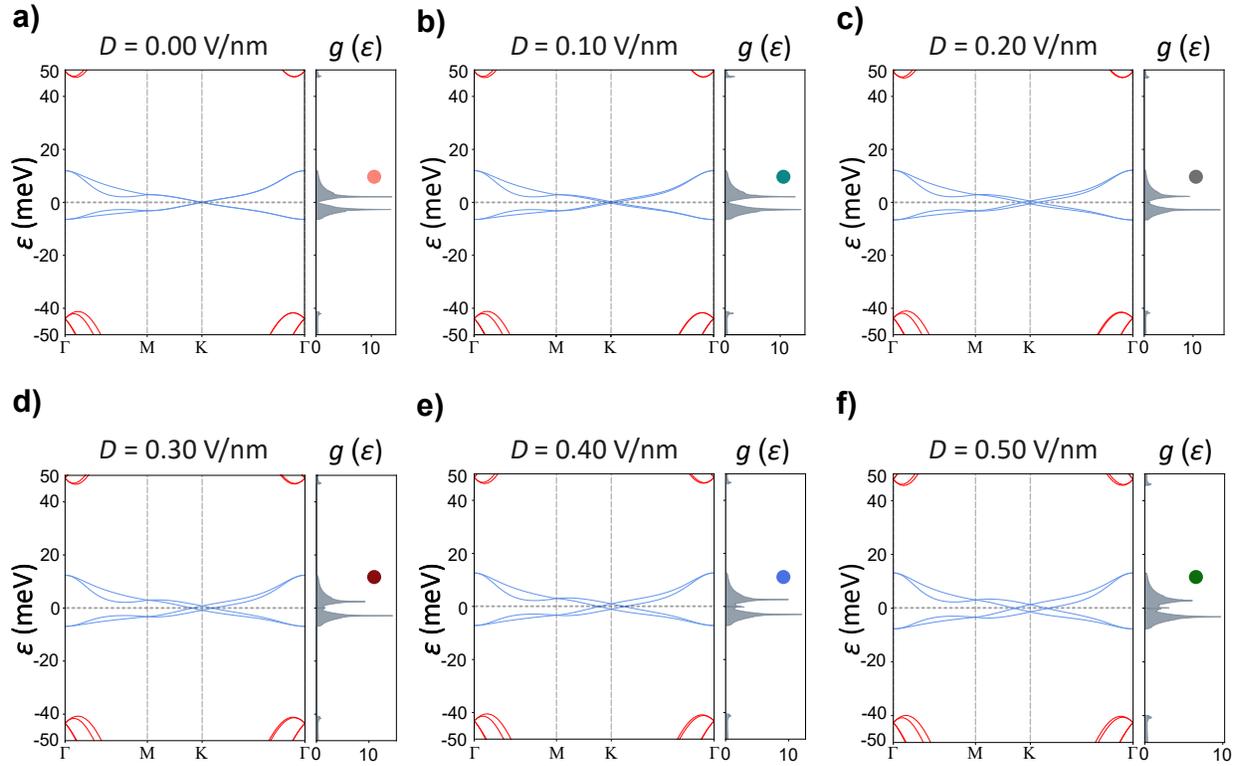

**SI-Fig. 18:** The calculated band structures using a tight-binding model on relaxed structures of twist angle ~ 0.95° tBLG for $D = 0.00, 0.10, 0.20, 0.30, 0.40$, and $0.50$ V/nm. The colors of the circles in the right panels of the figures correspond to the colors assigned to the DOS vs $\nu$ plots at different $D$ in SI-Fig. 19.

23

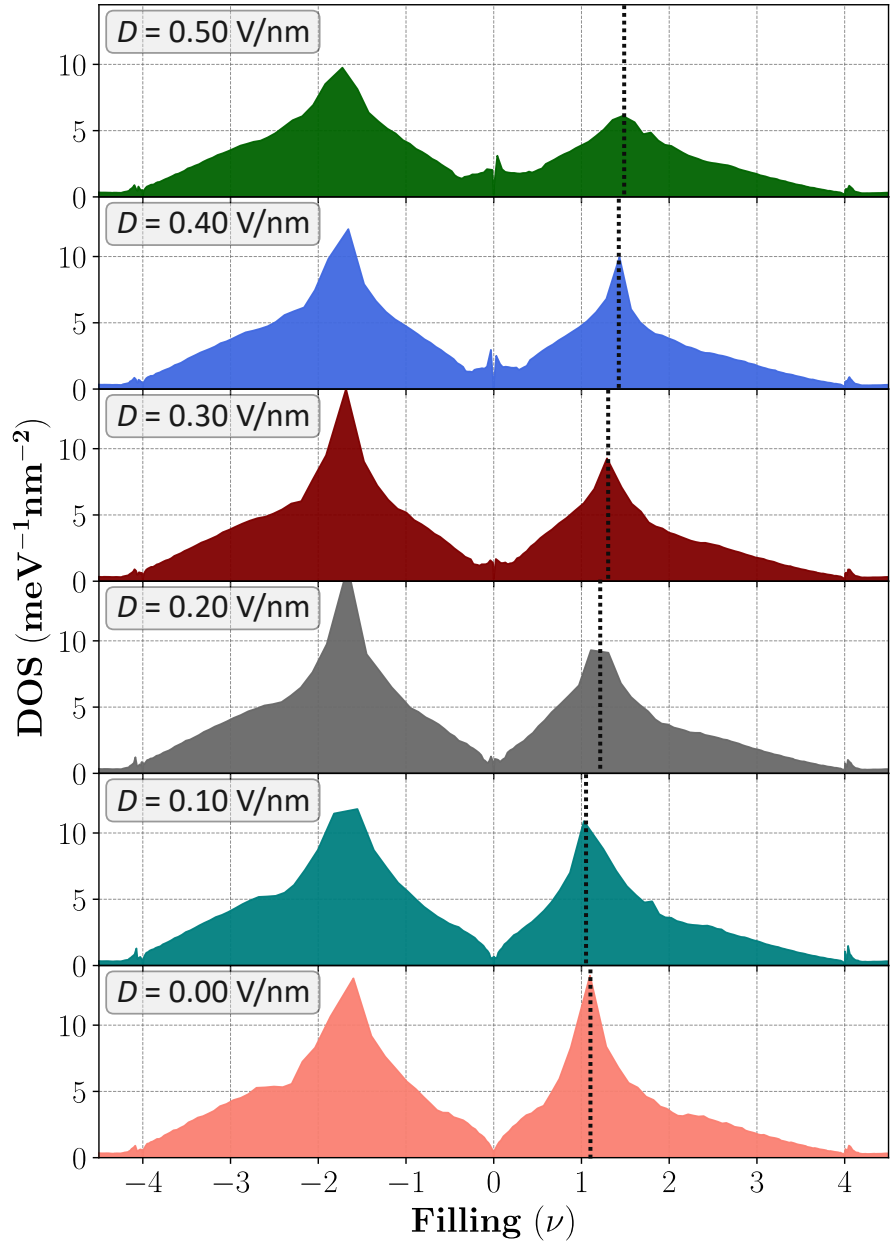

**SI-Fig. 19:** Density of states at the Fermi energy plotted as a function of filling ($\nu$) for different displacement fields.


\end{document}